\documentclass[a4paper]{JHEP3}

\usepackage[dvips]{graphicx}	
\usepackage{amsbsy}		
\usepackage{amsfonts}		
\usepackage{amsmath}		
\usepackage{amssymb}		
\usepackage{slashed}
\usepackage{cite}

\usepackage{psfrag}
\usepackage{subfigure}
\usepackage[latin1]{inputenc}
\usepackage[dvips]{graphicx} 
\usepackage{slashed}

\newcommand{\sd}{\mathrm{d}}

\newcommand{\Mtt}{M_{t\bar{t}}}
\newcommand{\ttbar}{t\bar{t}}
\newcommand{\muF}{\mu_\mathrm{F}}

\newcommand{\twobar}{\overline{\mathrm{II}}}
\newcommand{\mg}{m_{\tilde{g}}}
\newcommand{\mn}{m_{\tilde{\chi}^0_a}}
\newcommand{\sqt}{{\tilde{t}}}
\newcommand{\sqb}{{\tilde{b}}}
\newcommand*{\Ave}[1]{\mathinner{\left\langle{#1}\right\rangle}}

\title{New angles on top quark decay to a charged Higgs}
\author{David Eriksson, Gunnar Ingelman, Johan Rathsman, Oscar St\aa l$^*$\\

 High-Energy Physics, Dept.~of Nuclear and Particle Physics\\
 Uppsala University, P.\,O.\,Box 535, SE-751\,21 Uppsala,  Sweden\\ 
 $^*$Corresponding author. E-mail: \email{oscar.stal@tsl.uu.se}

}

\abstract{To properly discover a charged Higgs Boson ($H^\pm$) requires its spin and couplings to be determined. We investigate how to utilize $\ttbar$ spin correlations to analyze the $H^\pm$ couplings in the decay $t\to bH^+\to b\tau^+\nu_\tau$. Within the framework of a general Two-Higgs-Doublet Model, we obtain results on the spin analyzing coefficients for this decay and study in detail its spin phenomenology, focusing on the limits of large and small values for $\tan\beta$. Using a Monte Carlo approach to simulate full hadron-level events, we evaluate systematically how the $H^\pm\to\tau^\pm\nu_\tau$ decay mode can be used for spin analysis. The most promising observables are obtained from azimuthal angle correlations in the transverse rest frames of $t(\bar{t})$. This method is particularly useful for determining the coupling structure of $H^\pm$ in the large $\tan\beta$ limit, where differences from the SM are most significant.}


\begin{document}



\section{Introduction}
Finding a fundamental spin zero boson with electric charge would be a direct sign of physics beyond the Standard Model (SM). The existence of a charged Higgs boson pair ($H^\pm$) with these properties is predicted by Two-Higgs-Doublet Model (2HDM) extensions of the SM Higgs sector. The primary motivation for studying the 2HDM is supersymmetry, which requires an even number of Higgs doublets for cancellation of triangle anomalies. 
Charged Higgs searches at hadron colliders are divided into two regimes, separated by the dominant mode of production. When $H^\pm$ is heavy ($m_{H^+}\gtrsim m_t$), it is produced primarily through the $gg\to H^+\bar{t}b$ and $gb\to H^+\bar{t}$ processes \cite{Barnett&al:NPB:1987,Bawa&al:ZP:1989,Borzumati&al:PRD:1999,Miller&al:PRD:1999}. When, on the other hand, $H^\pm$ is light ($m_{H^+}<m_t-m_b$) and the decay $t\to bH^+$ \cite{Glashow&Jenkins:PLB:1987,Barger&Phillips:PLB:1987,Barger&Phillips:PRD:1989,Barger&Phillips:PRD:1990,Godbole&Roy:PRD:1991} opens up, this quickly becomes the dominant production mode.

The most stringent model-independent limit on the mass of $H^\pm$ from a direct search experiment comes from LEP: $m_{H^\pm}>79.3$~GeV \cite{ALEPH:PLB:2002} at $95\%$ CL, assuming only the decays $H^+\to c\bar{s}$ and $H^+\to\tau^+\nu_\tau$ are possible. Even tighter constraints on $m_{H^+}$ have later been derived using Tevatron data \cite{CDF:PRL:2006}, but these are not independent of the other 2HDM parameters. Neither are indirect constraints on $m_{H^+}$ obtained from $B$-physics observables.

The upcoming searches for $H^\pm$ planned by the LHC experiments will have good sensitivity to discover $H^\pm$ over a wide parameter range \cite{Denegri&al:2001,Assamagan&al:EPJ:2002,Biscarat&Dosil:ATLAS:2003,Baarmand&al:JPG:2006,Mohn:ATLAS:2007}, especially when $H^\pm$ is light. However, even if some candidate $H^\pm$ state was to be found, this discovery alone would not be enough to establish the validity of the Higgs mechanism as described by the 2HDM. To do this requires further that the spin, and the couplings, of this new particle be determined. Here we investigate one possibility to address this issue when $m_{H^+}<m_t-m_b$. In addition to providing a discovery channel, the $t\to bH^+$ decay mode could modify the ordinary V-A Lorentz structure of weak top decay significantly. As we will show, this fact can provide a handle on the spin and coupling structure of $H^\pm$ by making use of spin correlations.

Top quarks produced in pairs at hadron colliders constitute an interesting laboratory for observing spin
effects in high-energy physics. Since the timescale for weak top decay
$1/\Gamma_t$ is much shorter than the typical hadronization timescale
$1/\Lambda_\mathrm{QCD}$, the heavy quarks will decay before hadrons can form
\cite{Bigi&al:PLB:1986}. No hadronic effects will therefore obfuscate the
spin information. Unlike the case for the lighter quarks, this fact
allows for reliable perturbative calculations of the relevant spin observables.
Furthermore, since the charged current weak interaction violates parity
maximally, the decay self-analyzes the spin of the top quark. This means the full spin information will be imprinted in the angular
distributions of the different decay products.

In order for the angular information to be useful for investigating the couplings involved, a method to determine the spin projection of the decaying top quark is required. At a hadron collider, this can be achieved by exploiting \emph{correlations} between the top quark spins. By using the decay information from one side of a $\ttbar$ event, it is possible to determine statistically the polarization of the other top. 
Spin correlations in top quark pair production and decay have been extensively discussed within the
Standard Model (SM) for hadron-hadron colliders
\cite{Kuhn:PLB:1984,Barger&al:IJMP:1989,Hara:PTP:1991,Arens&Sehgal:PLB:1993,
Chang&al:CJP:1995,Mahlon&Parke:PRD:1996,
Stelzer&Willenbrock:PLB:1996,Chang&al:PRL:1996,Brandenburg:PLB:1996,Mahlon&Parke:PLB:1997,
Bernreuther&al:PRL:2001,Bernreuther&al:NPB:2004,Frixione&al:JHEP:2007}, and for
$e^+ e^-$ experiments
\cite{Kane&al:PRD:1991,Arens&Sehgal:NPB:1993,Schmidt:PRD:1996,
Parke&Shadmi:PLB:1996, Brandenburg&al:PRD:1998,Liu&al:PLB:1999}. Utilizing the
top spin information to study physics beyond the Standard Model was considered
in the context of anomalous $Wtb$-couplings
\cite{Kane&al:PRD:1991,delAguila&al:PRD:2002,AguilarSaavedra&al:EPJC:2006}, for
probing extended Higgs sectors and effects of CP-violation
\cite{Schmidt&Peskin:PRL:1992,Bernreuther&al:NPB:1992, Brandenburg&Ma:PLB:1993,
Bernreuther&Brandenburg:PLB:1993,
Bernreuther&Brandenburg:PRD:1994,Korner&Mauser:hep-ph:2002} , and within
theories with extra dimensions \cite{Arai&al:PRD:2004,Arai&al:PRD:2007}.


We study the spin phenomenology of a light scalar sector in full generality, ignoring indirect constraints on the 2HDM. We will however restrict the
mass to $m_{H^+}\gtrsim m_W$, as required by the non-observation of $H^\pm$ in
direct search experiments. Due to the, in the context of spin observables, relatively limited sample of $t\bar{t}$ events
available from the Tevatron, our main focus will be on prospects for
observations at the LHC. We try to comment on issues that are of relevance
also for the analysis of Tevatron data.

The organization of this paper is as follows. First, in Section~\ref{Sect:Production}, we discuss $\ttbar$ pair
production at hadron colliders, and how spin correlations come about in this
process. Then in
Section~\ref{Sect:Decay} we introduce the phenomenological model, followed by
a brief review of the relevant theory for polarized top quark decay, and
results on the spin analyzing efficiencies in models where the top
quark can decay through a charged Higgs boson. Section~\ref{Sect:Simulations}
describes a Monte Carlo simulation study of these effects, and discusses possible
experimental observables. Finally, Section~\ref{Sect:Conclusions} contains a
summary and the conclusions of this work.

\section{Top Quark Pair Production at Hadron Colliders}
\label{Sect:Production}
We will adopt the latest combined value for
the top mass $m_t=170.9\pm 1.8$ GeV \cite{Topmass:2007}.
Pair production of top quarks occurs in leading order QCD both through
$q\bar{q}$ annihilation via an $s$-channel gluon, and through gluon fusion for
which $s$-, $t$- and $u$-channel exchanges are possible. The total hadronic
cross section has long been known to NLO accuracy
\cite{Nason&al:NPB:1988,Nason&al:NPB:1989}. With CTEQ6 parton distributions \cite{CTEQ6.1:JHEP:2002} and a common choice of scales $\mu_\mathrm{R}=\mu_\mathrm{F}=m_t$, we obtain $\sigma(pp\to t\bar{t}+X)\simeq900$~pb at $\sqrt{s}=14$ TeV using the NLO MC generator POWHEG \cite{Frixione&al:preprint:2007}. One year of LHC running at low luminosity,
corresponding to $\int\mathcal{L}=10$ fb$^{-1}$, will therefore produce of order
$10^7$ $\ttbar$ events.

Individually unpolarized top quarks are still produced with strong correlations
between their spin projections in a suitable basis. The nature and magnitude of
these correlations depend on the partonic center of mass (CM) energy. Following
\cite{Stelzer&Willenbrock:PLB:1996} we define the production correlation, as a
function of the invariant mass $\Mtt$ of the top pair, to be
\begin{equation}
\label{Eq:Cij}
\hat{C}_{ij}(M_{t\bar{t}}^2) =
\frac{\hat{\sigma}_{ij}(t_\uparrow\bar{t}_\uparrow+t_\downarrow\bar{t}
_\downarrow)-\hat{\sigma}_{ij}(t_\downarrow\bar{t}_\uparrow+t_\uparrow\bar{t}
_\downarrow)}{\hat{\sigma}_{ij}(t_\uparrow\bar{t}_\uparrow+t_\downarrow\bar{t}
_\downarrow)+\hat{\sigma}_{ij}(t_\downarrow\bar{t}_\uparrow+t_\uparrow\bar{t}
_\downarrow)}
\end{equation}
for the partonic subprocess involving initial state partons $(i,j)$. Arrows indicate
the spin projection on the chosen spin quantization axes (which may be different
for $t$ and $\bar{t}$). We choose here to work exclusively in the \emph{helicity
basis}, in which the spin is quantized along the momentum directions of the
$t(\bar{t})$ in the partonic CM frame. In this basis, the notation $(R,L)$ is
sometimes used interchangeably with $(\uparrow, \downarrow)$ to denote the two
spin projections.

Near threshold, the $t\bar{t}$ are always produced in an S-wave state.
For production dominated by $q\bar{q}\to t\bar{t}$ through an $s$-channel gluon,
the overall angular momentum state will therefore be $^3\mathrm{S}_1$. Out of
the three states composing the triplet, two correspond to opposite helicities
for the two top quarks, whereas one state gives equal helicities. The combined
correlation according to (\ref{Eq:Cij}) is thus
$\hat{C}_{q\bar{q}}(4m^2_{t})=-1/3$ at threshold.
When instead $gg\to t \bar{t}$ production dominates, which is the
case for LHC energies, the situation at threshold is reversed. The top pair is now produced
in a singlet $^1\mathrm{S}_0$ configuration, since the initial state gluons do
not populate $J=1$ states \cite{Hara:PTP:1991}. In
this case the $t\bar{t}$ must always come with the same helicities, which means that
$\hat{C}_{gg}(4m^2_{t})=1$.
Finally, in the ultra-relativistic limit
($M^2_{t\bar{t}}\gg 4m^2_t$), helicity conservation requires the $\ttbar$ to have
opposite helicities independent of partonic subprocess. Hence the correlation $\hat{C}_{ij}(M_{t\bar{t}}^2)\to-1$ for all $(i,j)$
in the high energy limit. 

\begin{figure}
\begin{centering}
   \includegraphics[width=0.5\columnwidth,keepaspectratio]{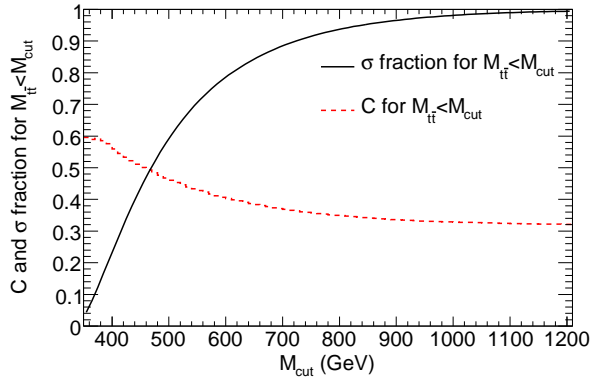}
\caption{The dashed red curve shows the resulting value of $\mathcal{C}$ in the helicity basis, at the LHC, when a cut $M_{\ttbar}<M_\mathrm{cut}$ has been applied. The black curve shows the fraction of the total cross section which survives the cut.}
\label{Fig:inteff}
\end{centering}
\end{figure}
In general, the total statistical correlation $\mathcal{C}$ in a sample of
$\ttbar$ events is obtained from averaging over the invariant mass of the top
pair, while parton distribution functions determine the relative contributions
of the competing production processes:
\begin{equation}
\mathcal{C}(s) = \frac{1}{\sigma_{\ttbar}}\sum_{i,j=\{q,\bar{q},g\}}\int \mathrm{d}x_1 \mathrm{d}x_2
\left[\hat{\sigma}_{ij}(t_\uparrow\bar{t}_\uparrow+t_\downarrow\bar{t}
_\downarrow)-\hat{\sigma}_{ij}(t_\downarrow\bar{t}_\uparrow+t_\uparrow\bar{t}
_\downarrow)\right]f_i(x_1,\muF^2)f_j(x_2,
\muF^2) .
\end{equation}
The $\hat{C}_{ij}$ for all partonic
subprocesses have previously been calculated to NLO in QCD. From these one obtains $\mathcal{C}=0.326$ \cite{Bernreuther&al:NPB:2004} in the helicity basis
for $pp$ collisions at $\sqrt{s}=14$ TeV. 
   The residual uncertainty in this
number from PDF and scale choices is of order one percent, a value similar to the difference from the LO calculation which gives $\mathcal{C}=0.319$.
For Tevatron run-II conditions ($p\bar{p}$ collisions at $\sqrt{s}=1.96$ TeV)
the same helicity basis correlation becomes $\mathcal{C}=-0.352$. At Tevatron
energies, where $q\bar{q}$ annihilation totally dominates $\ttbar$ production,
there exist also more efficient bases for spin quantization in which
correlations as large as $\mathcal{C}=0.8$ can be obtained \cite{Mahlon&Parke:PLB:1997}.

As suggested by the discussion above, it can be beneficial to introduce
an experimental cut on $\Mtt$ to increase the spin-purity of the $\ttbar$ sample at the
cost of decreased efficiency \cite{Mahlon&Parke:PRD:1996}. To increase the component of like-sign helicities in the gluon sample at the LHC requires a cut on the \emph{maximum} $M_{\ttbar}$. The effect of such a cut on the correlation parameter $\mathcal{C}$ is shown in Figure~\ref{Fig:inteff}. To illustrate the trade-off between spin-purity and efficiency, Figure~\ref{Fig:inteff} also presents the fraction of the total cross section which passes a cut on the maximum $M_{\ttbar}$. As we have already indicated, $\ttbar$ statistics will not be the limiting factor at the LHC. It is therefore good to keep in mind that $\mathcal{C}$ can be increased using this technique, although we do not make explicit use of this fact here.

\section{Spin Information in Top Quark Decay}
\label{Sect:Decay}
In the SM, the top quark decays almost exclusively via the charged current V-A
vertex
\begin{equation}
\label{Eq:Vff}
\mathcal{L}_{Wtb}=\frac{g_W}{\sqrt{2}}V_{tb}
W_\mu^+\bar{t}\gamma^\mu\frac{1-\gamma^5}{2}b+\mathrm{h.c.},
\end{equation}
where $V_{tb}\sim 1$ is the appropriate element of the CKM matrix. In addition
to the Standard Model decay, the possibility exists that the top quark decays
anomalously. The decay could then contain a small V+A component, or it could be
mediated by additional bosons of different spin. The study of spin correlations opens a window on
both these possibilities. Our aim here is to explore the latter case, allowing for an
extended scalar sector. 

\subsection{Charged Higgs Model}
Introducing a charged scalar pair $H^\pm$, their interactions with fermions
are parametrized by an effective Lagrangian density
\begin{equation}
\begin{aligned}
\label{Eq:HiggsL}
\mathcal{L}_H=&\frac{g_W}{2\sqrt{2}m_W}
\sum_{\substack{\{u,c,t\}\\\{d,s,b\}}}V_{ud}\Biggl\{
H^+\bar{u}\Bigl[A\left(1-\gamma_5\right)+B\left(1+\gamma_5\right)\Bigr]d+H^-\bar
{d}\Bigl[B^*\left(1-\gamma_5\right)+A^*\left(1+\gamma_5\right)\Bigr]u\Biggr\} 
\\
&+\frac{g_W}{2\sqrt{2}m_W}\sum_{\{e,\mu,\tau\}}\Bigl[H^+C\bar{\nu}
_l\left(1+\gamma_5\right)l+H^-C^*\bar{l}\left(1-\gamma_5\right)\nu_l\Bigr] .
\end{aligned}
\end{equation}
The $A$, $B$, $C$, and their complex conjugates, are in principle free parameters
determining the Lorentz structure of the couplings. Note that this model does not assign
definite parity to the $H^\pm$ unless $B=\pm A$. For $AB=0$ parity is violated
maximally. Assuming CP-invariance of the scalar sector, the coupling parameters
can all be taken as real numbers. 

A model such as (\ref{Eq:HiggsL}) occurs, for example, as the charged
Higgs-fermion sector of a two Higgs Doublet Model (2HDM). Here, the SM Higgs
sector is augmented with another complex Higgs doublet, resulting in two charge
conjugate ($H^\pm$) and three neutral ($h,H,A$) states occurring as physical
bosons. To ensure the augmented SM does not allow for tree-level FCNC's, certain
restrictions apply on how to couple the extended Higgs sector to the fermions.
For our purposes, it suffices to say that two options are generally considered: In the
so-called type I model [2HDM (I)], only one doublet is coupled directly to the
fermions. In the type II model [2HDM (II)], one doublet is coupled only to
up-type fermions, whereas the other doublet couples only to down-type fermions.
\begin{table}
\centering
 \caption{CP-invariant couplings of charged Higgs bosons to fermions in 2HDM
(I), 2HDM (II) and SUSY-improved type II model ($\twobar$).}.         
\begin{tabular}{cccc}
   \hline                
         Coupling  & 2HDM (I) & 2HDM (II) & 2HDM ($\overline{\mathrm{II}}$) \\
         \hline
          & & & \\[-6pt]        
         $A$  & $m_u\cot\beta$ & $m_u\cot\beta$ &
$m_u\cot\beta\left[1-\epsilon'_t\tan\beta\right]$\\[5pt]
         $B$  & $-m_d\cot\beta$ & $m_d\tan\beta$ &
$\frac{m_d\tan\beta}{1+\epsilon_b\tan\beta}$ \\[5pt]
         $C$  & $m_l\cot\beta$ & $m_l\tan\beta$ & $m_l\tan\beta$\\[5pt]
         \hline
\end{tabular}
\label{tab:ABC}
\end{table}
The number of independent parameters in the Higgs sector is thereby restricted
to two at leading order. We will adopt for these the ratio
$\tan\beta=v_2/v_1$ of the two doublets vacuum expectation values, and the
charged Higgs mass $m_{H^+}$. For the two model types, the charged Higgs-fermion couplings are then given in Table~\ref{tab:ABC}. 
Mass parameters appearing in the couplings should be evaluated at a scale
$Q^2=m_{H^+}^2$, using the
$\overline{\mathrm{MS}}$ masses $\overline{m}_q(Q^2)$ to ensure proper resummation of large logarithmic QCD vertex corrections
\cite{Carena:NPB:2000,Braaten&Leveille:PRD:1980,Drees&Hikasa:PLB:1990}. Since the $H^\pm$ couples proportionally to the fermion mass, we will only be concerned with third generation particles in the following.

As discussed in the introduction, one possible extension of the Standard Model
where a 2HDM occurs naturally is the Minimal Supersymmetric Standard Model (MSSM). The MSSM contains a 2HDM (II), but since the supersymmetry introduces additional particles, the two-parameter picture of the 2HDM (II) works only as an effective tree-level description. It has been shown \cite{Carena:NPB:2000,Degrassi:JHEP:2000}, that quantum corrections due to
SUSY-QCD loops can be quite sizable for large values of $\tan\beta$. This holds even in
the decoupling limit when all SUSY masses are taken to infinity. How these
corrections enter into the effective $H^\pm$ couplings can be seen from
the third column of Table~\ref{tab:ABC}. We will call the 2HDM which includes the $\tan\beta$ enhanced SUSY
corrections the "modified type II", or simply $\twobar$. The corrections are of two types:
first the so-called $\epsilon_b$ correction to the relation between the
bottom quark mass $m_b$ and the bottom Yukawa coupling $y_b$. It is caused
by gluino-sbottom and chargino-stop loops. At one-loop, the dominant contributions to this correction are given by
\cite{Hall&al:PRD:1994, Degrassi:JHEP:2000}
\begin{equation}
\epsilon_b=-\frac{2\alpha_s}{3\pi}\frac{\mu}{\mg}H_2\Biggl(\frac{m_{\tilde{b}_1}
}{\mg},\frac{m_{\tilde{b}_2}}{\mg}\Biggr)-\frac{y_t^2}{16\pi^2}\tilde{U}_{a2}
\frac{A_t}{m_{\tilde{\chi}_a^+}}H_2\Biggl(\frac{m_{\tilde{t}_1}}{m_{\tilde{\chi}
_a^+}},\frac{m_{\tilde{t}_2}}{m_{\tilde{\chi}_a^+}}\Biggr)\tilde{V}_{a2},
\end{equation}
which introduces a dependence on the trilinear coupling $A_t$, the top Yukawa
coupling $y_t$, and the $\mu$ parameter from the superpotential -- in addition to the dependence on several of the sparticle masses.
The real matrices $\tilde{U}$ and $\tilde{V}$ diagonalize the chargino
mass matrix. 
The function $H_2$ is given by
\begin{equation}
H_2(x,y)=\frac{x\ln x}{(1-x)(x-y)}+\frac{y\ln y}{(1-y)(y-x)}.
\end{equation}
In the limit when all SUSY parameters and sparticle masses are of similar scale
$M_\mathrm{SUSY}$ one obtains $|\epsilon_b|\simeq
\alpha_s(Q=M_\mathrm{SUSY})/(3\pi)\sim 10^{-2}$.
The sign of $\epsilon_b$ is determined by the sign of $\mu$. 

The second
contribution which modifies the $H^\pm$ couplings is \cite{Degrassi:JHEP:2000}
\begin{equation}
\begin{aligned}
\epsilon'_t=&-\frac{2\alpha_s}{3\pi}\frac{\mu}{\mg}
\left[c^2_\sqt c^2_\sqb
H_2\Biggl(\frac{m_{\sqt_2}}{\mg},\frac{m_{\sqb_1}}{\mg}\Biggr)+
c^2_\sqt s^2_\sqb
H_2\Biggl(\frac{m_{\sqt_2}}{\mg},\frac{m_{\sqb_2}}{\mg}\Biggr)\right.\\
&\left.\quad\quad\quad\quad\quad+s^2_\sqt c^2_\sqb
H_2\Biggl(\frac{m_{\sqt_1}}{\mg},\frac{m_{\sqb_1}}{\mg}\Biggr)+
s^2_\sqt s^2_\sqb
H_2\Biggl(\frac{m_{\sqt_1}}{\mg},\frac{m_{\sqb_2}}{\mg}\Biggr)\right]-\\
&-\frac{y_b^2}{16\pi^2}N_{4a}^*\frac{A_b}{\mn}\left[c^2_\sqt c^2_\sqb
H_2\Biggl(\frac{m_{\sqt_1}}{\mn},\frac{m_{\sqb_2}}{\mn}\Biggr)+
c^2_\sqt s^2_\sqb
H_2\Biggl(\frac{m_{\sqt_1}}{\mn},\frac{m_{\sqb_1}}{\mn}\Biggr)\right.\\
&\left.\quad\quad\quad\quad\quad+s^2_\sqt c^2_\sqb
H_2\Biggl(\frac{m_{\sqt_2}}{\mn},\frac{m_{\sqb_2}}{\mn}\Biggr)+
s^2_\sqt s^2_\sqb
H_2\Biggl(\frac{m_{\sqt_2}}{\mn},\frac{m_{\sqb_1}}{\mn}\Biggr)\right]N_{a3},
\end{aligned}
\end{equation}
where the matrix $N$ diagonalizes the neutralino mass matrix, $s_{\tilde{q}}=\sin\theta_{\tilde{q}}$ and $c_{\tilde{q}}=\cos\theta_{\tilde{q}}$ for the squark mixing angles $\theta_{\tilde{q}}$. The squark mass
eigenstates are given by
$\tilde{q}_1=c_{\tilde{q}}\tilde{q}_\mathrm{L}+s_{\tilde{q}}\tilde{q}_\mathrm{R}
$ and
$\tilde{q}_2=-s_{\tilde{q}}\tilde{q}_\mathrm{L}+c_{\tilde{q}}\tilde{q}_\mathrm{R
}$, with $m_{\tilde{q}_1}>m_{\tilde{q}_2}$. We note that $\epsilon_t'$ is
numerically similar to $\epsilon_b$ in the case with a common scale for the SUSY
parameters. In Section~\ref{Sect:2HDMnum}, we will return to these SUSY corrections when discussing numerical results for the 2HDM. As we will show, it turns out that their effects on $\ttbar$ spin correlation observables are small.  

\subsection{Top Quark Decay with Polarization}
\begin{figure}
\begin{centering}

\subfigure{
   \label{Fig:tdecayW}
   \includegraphics[width=0.45\columnwidth,keepaspectratio]{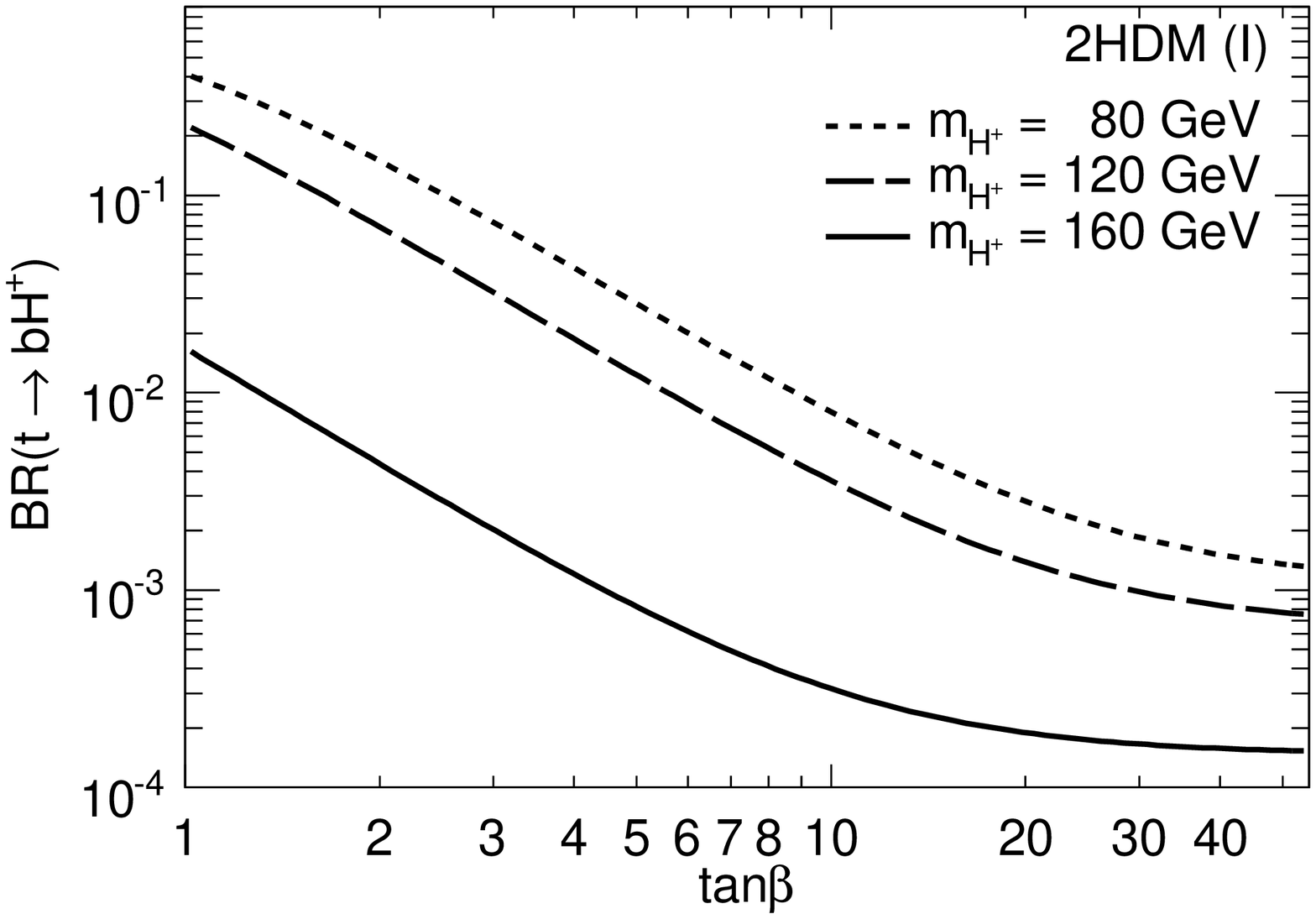}

}
\subfigure{
   \label{Fig:tdecayH}
   \includegraphics[width=0.45\columnwidth,keepaspectratio]{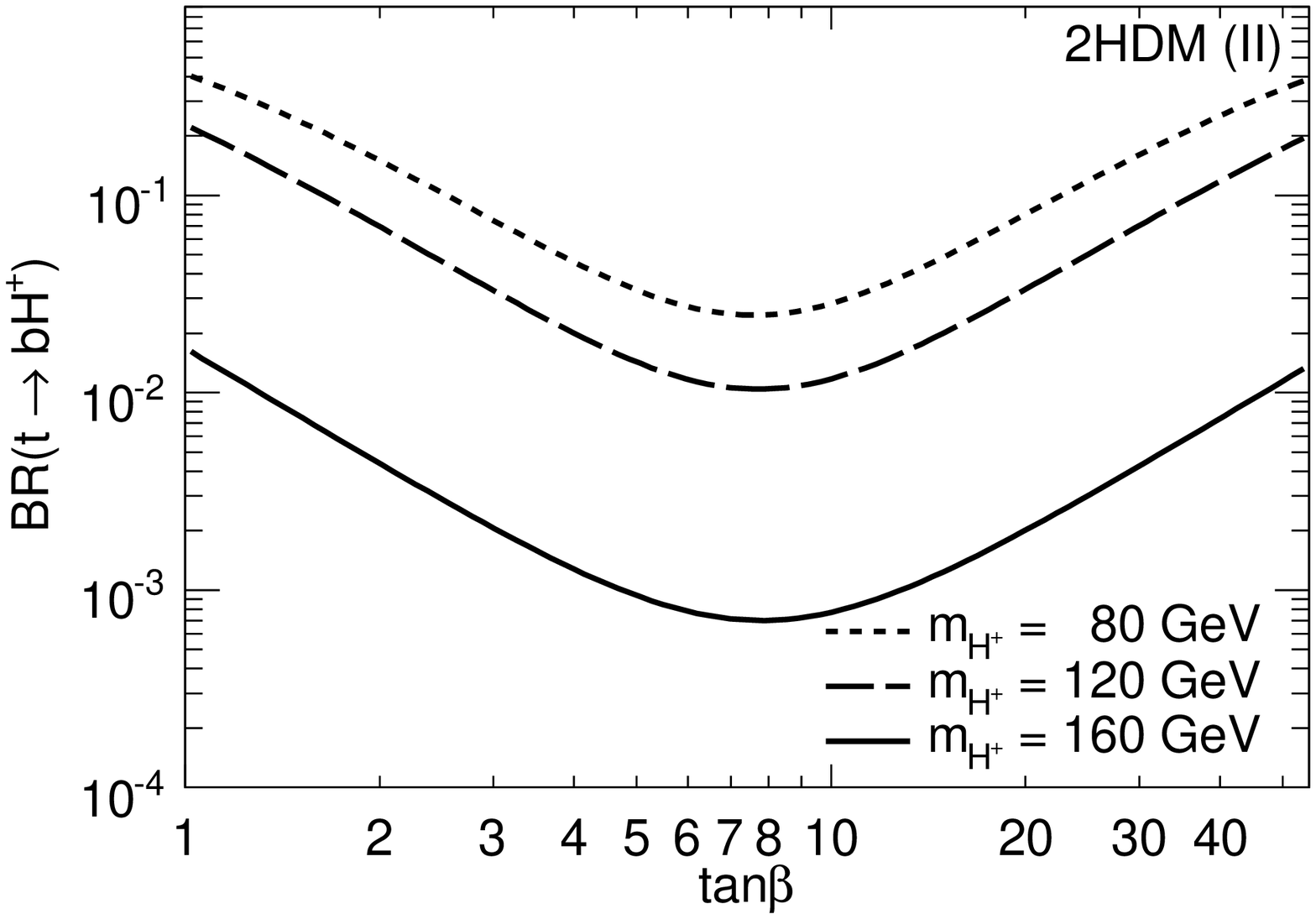}
}
\caption{Branching fractions for $t\to bH^+$ in 2HDM (I) (left) and 2HDM (II) (right), as a function of $\tan\beta$. The contours show $m_{H^+}=80$ GeV (short-dashed), $m_{H^+}=120$ GeV (long-dashed), and $m_{H^+}=160$ GeV (solid).}
\label{Fig:BR}
\end{centering}
\end{figure}
Assuming that the full width $\Gamma_t$ of the top quark, including the scalar
decay mode, is still very small ($\Gamma_t/m_t\lesssim 0.01$) we use the narrow
width approximation to factorize the production from the decay of the heavy
quarks. The branching fractions for $t\to bH^+$ are shown in Figure~\ref{Fig:BR} for 2HDM type (I) and (II). It is clear that type (II) is interesting both for small and large $\tan\beta$ values, whereas the 2HDM (I) only allows a significant $\mathcal{BR}(t\to bH^+)$ for small $\tan\beta$. In the following we will mostly be concerned with the type (II) model.

We treat the decaying $\ttbar$ as independent decays through well-defined channels without interference effects. Strictly speaking, a more complete formalism involving off-diagonal $W^\pm/H^\pm$ propagator elements could be used when $m_{H^+}\simeq m_{W}$. Ignoring such complications, the full structure of density matrices in the $2\to 6$ matrix element becomes 
\begin{equation}
|\mathcal{M}(2\to
6)|^2=[R_{\lambda\lambda'\kappa\kappa'}(2\to \ttbar)\otimes
\rho^i_{\lambda\lambda'}(t\to 3)\otimes \rho^j_{\kappa\kappa'}(\bar{t} \to 3)].
\end{equation}
$R$ is here the fully helicity-dependent spin density matrix for $\ttbar$
production. The $\rho$($\bar{\rho}$) are decay density matrices of
$t$($\bar{t}$), where $i,j$ label the available decay channels and
$\lambda,\lambda'$ ($\kappa, \kappa'$) are helicity indices for the
$t$($\bar{t}$). 

\begin{figure}
\begin{centering}

\subfigure{
   \label{Fig:tdecayW}
   \includegraphics[width=0.45\columnwidth,keepaspectratio]{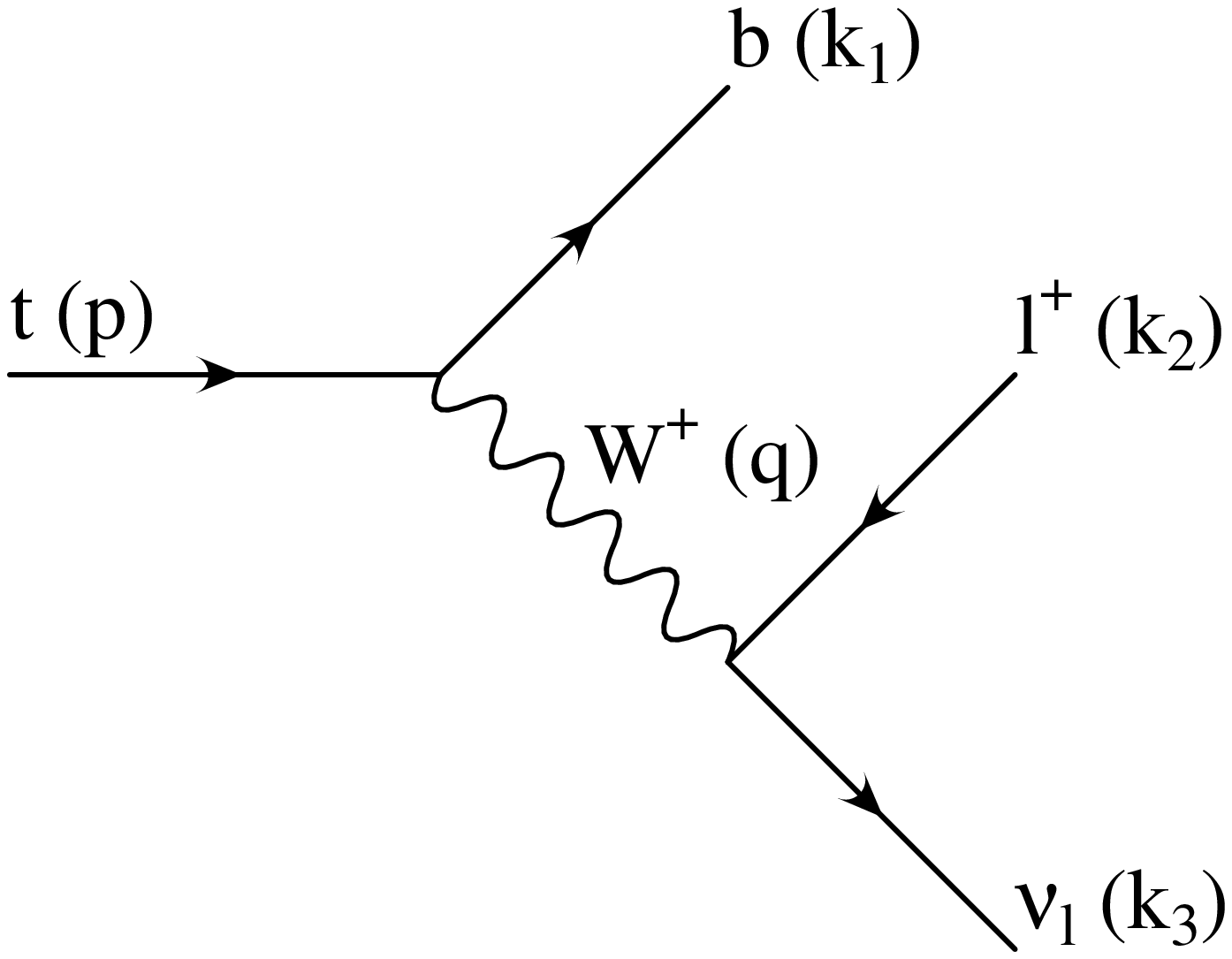}

}
\subfigure{
   \label{Fig:tdecayH}
   \includegraphics[width=0.45\columnwidth,keepaspectratio]{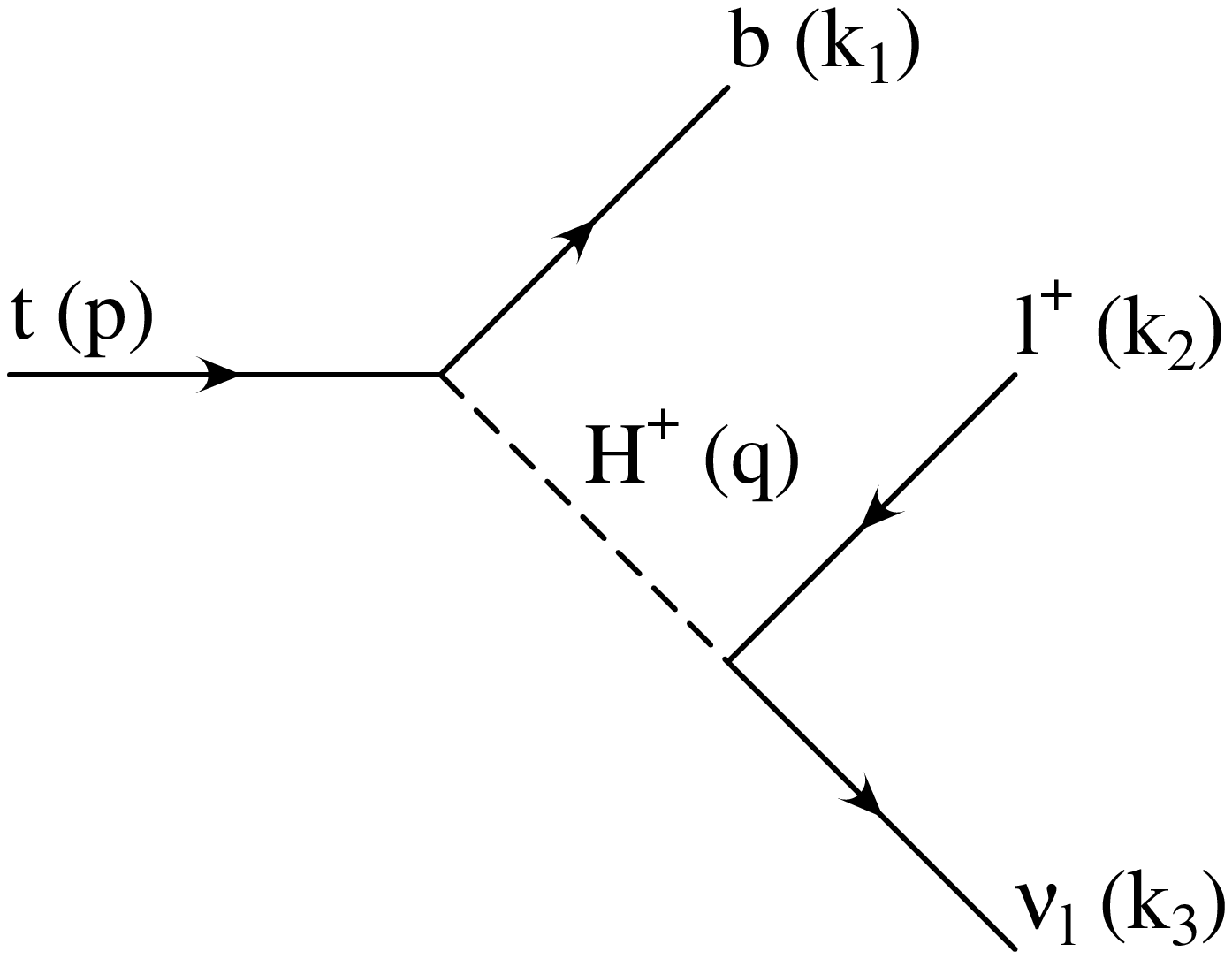}
}
\caption{Notation for momenta in semi-leptonic top quark decay mediated either
by $W^+$ or $H^+$. In hadronic decay of the boson, we use the same notation while making the
replacements of weak isospin partners: $l^+\to \bar{d}$, and $\nu_l \to u$.}
\label{Fig:tdecay}
\end{centering}
\end{figure}
To obtain the decay density matrix for a given channel, we use the techniques
described in Appendix \ref{App:Spin}. With momenta defined in
Figure~\ref{Fig:tdecay}, the leading order decay density matrix elements for
semi-leptonic weak decay of the top quark are given by
\begin{equation}
\label{Eq:rhoW}
\rho_{\lambda\lambda'}^W=|\mathcal{M}_{\lambda\lambda'}(t\to bW^+\to
bl^+\nu_l)|^2=\frac{2g_W^4|V_{tb}|^2(p\cdot k_2)(k_1 \cdot
k_3)}{(q^2-m_W^2)^2+m_W^2\Gamma_W^2}\left[\delta_{\lambda\lambda'}+\hat{k}
_2^a\sigma^a_{\lambda\lambda'} \right]
\end{equation}
when the spins of all outgoing particles are summed over. The unit 3-vector
$\hat{k}_2$ is given in the rest frame of the decaying quark. For hadronic decay
of the $W$ boson, the matrix elements are exactly the same if a)~all final state masses
are neglected and b)~the leptons are replaced by their quark counterparts in
terms of weak isospin.
CP-invariance of the decay ensures that $|\mathcal{M}_{\lambda\lambda'}(t\to
bl^+\nu_l)|^2=|\mathcal{M}_{\lambda'\lambda}(\bar{t}\to
\bar{b}l^-\bar{\nu}_l)|^2$. 

Reckon similarly the elements of the decay density matrix when the decay is
mediated by a charged scalar as defined by the model (\ref{Eq:HiggsL}). In this
case, the elements become
\begin{equation}
\begin{aligned}
\label{Eq:rhoH}
\rho_{\lambda\lambda'}^H=&|\mathcal{M}_{\lambda\lambda'}(t\to bH^+\to
bl^+\nu_l)|^2=\frac{g_W^4|V_{tb}|^2(p\cdot k_1)(k_2\cdot k_3)}{(q^2-m_{H^+}^2)^2+m_{
H^+}^2\Gamma_{H^+}^2}\frac{C^2(A^2+B^2)}{2m_W^4} \\
&\times\left(1+\frac{AB}{A^2+B^2}\frac{4\delta}{1-\xi}\right)\left[
\delta_{\lambda\lambda'}-\frac{A^2-B^2}{A^2+B^2}\left(1+\frac{AB}{A^2+B^2}\frac{
4\delta}{1-\xi}\right)^{-1}\hat{k}_1^a\sigma^a_{\lambda\lambda'} \right].
\end{aligned}
\end{equation}
 Here the notation $\xi=m_{H^+}^2/m_t^2$ and $\delta = m_b/m_t$ is used. Let us
also introduce a convenient short-hand
\begin{equation}
f(\xi,A,B)=\left(1+\frac{AB}{A^2+B^2}\frac{4\delta}{1-\xi}\right)^{-1}
\end{equation}
for the threshold factor. This function has the general properties $f(\xi,A,B)=1$
for $m_{H^+}\ll m_t$, and $f(\xi,A,B)\to 0$ for $\xi\to 1$, unless $AB=0$.

\subsection{$W$ Boson helicity}
The perhaps most direct test of V-A theory in top quark decay is offered by
examining the polarization states of the $W$ boson mediating the decay. Due to the large Yukawa
coupling $y_t\sim 1$, a fraction $m_t^2/(m_t^2+2m_W^2)\simeq 0.69$ of the 
$W$ bosons are expected to be longitudinally polarized, while the remainder
carries a left-handed helicity in the $t$ rest frame. Uncertainties in these
 numbers from higher
order corrections, including virtual 2HDM and SUSY effects, are under control at the
$1\%$ level
\cite{Grzadkowski&Hollik:NPB:1992,Cao&al:PRD:2003}.

When the $W$ decays further, the angular dependencies of the decay products on
the different helicity states are given by the Wigner $d$-functions for the
spin~$1$ representation. Combining this knowledge with the polarized matrix
element for $t\to bW^+$, the normalized lepton angular distribution in leptonic decay
of the $W$ is given by
\begin{equation}
\label{Eq:ldistW}
\frac{1}{N}\frac{\sd N(W\to l \nu_l)}{\sd
\cos\theta_l^*}=\frac{3}{4(m_t^2+2M_W^2)}\left[
m_t^2\sin^2\theta_l^*+M_W^2(1-\cos\theta_l^*)^2\right],
\end{equation}
where $\theta^*_l$ is defined in the $W$ rest system as the angle of the
lepton momentum to the $W$ helicity axis. Using the fact that, in the rest frame
of the decaying top, the recoiling $b$ quark has its momentum anti-parallel to
that of the $W$, the lepton helicity angle $\cos\theta_l^*$ can be determined by
the invariant product \cite{ESW:QCD@Colliders:1996}
\begin{equation}
\cos\theta_l^*=\frac{k_1\cdot(k_2-k_3)}{k_1\cdot(k_2+k_3)}
\end{equation}
if the $b$ mass is neglected. Assuming further that the decay is mediated
through an on-shell $W$, the approximate expression
\begin{equation}
\label{Eq:costheta}
\cos\theta_l^*\simeq\frac{4k_1\cdot k_2}{m_t^2-m_W^2}-1
\end{equation}
can be obtained from the kinematics of the decay. The form
(\ref{Eq:costheta}) is experimentally advantageous since no knowledge of the
neutrino momenta is required to determine $\cos\theta^*_l$. Being only an
approximate on-shell relation, the values obtained using this expression may in
reality be such that $|\cos\theta^*_l|> 1$ for some events.

In the decay of a charged Higgs boson, the decay products should be
isotropically distributed in $\cos\theta^*$. This offers a
clear signature for a new charged boson to have spin~$0$.
However, even with a large branching ratio $t\to bH^+$, this
appreciable difference would not contribute much to measurements of the
distribution (\ref{Eq:ldistW}) using electrons or muons, simply because in the
$\tan\beta$ regions of interest, $\mathcal{BR}(H^\pm\to\tau^\pm\nu_\tau)\simeq 1$ in the 2HDM.\footnote{A small contamination from $H^\pm\to\tau^\pm\nu_\tau\to l^\pm\nu_l\nu_\tau\bar{\nu}_\tau$ would of course be present also in the lepton samples in these cases.} If, at the
LHC, evidence starts to gather in favor of a light $H^\pm$, it would
therefore be interesting to study the angular distribution of $\tau$~leptons exclusively
using the hadronic $\tau$ decay. Experimentally this presents a formidable task,
since the presence of two final state neutrinos in the $\tau$ channel introduces
ambiguities in the reconstruction of the $\tau$ momentum. Furthermore, when
$m_{H^+}>m_W$, the kinematic assumptions behind Equation (\ref{Eq:costheta}) are
no longer valid. It is then natural to exploit these kinematic differences fully
and treat the two decays separately also in the angular analysis in order to establish
the spin~$0$ nature of the presumptive $H^\pm$.

\begin{figure}
\begin{centering}
\includegraphics[width=0.5\columnwidth, keepaspectratio]{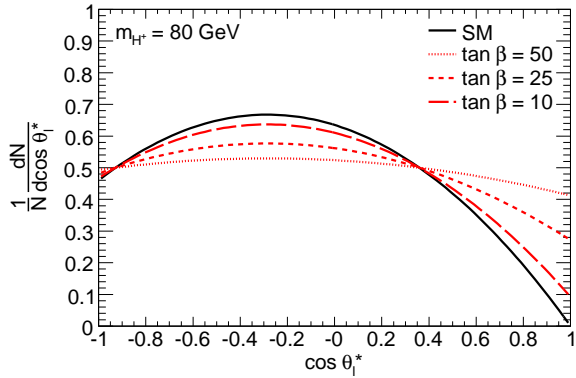}
\caption{Angular distribution of $\tau$ leptons in the rest system of the boson
mediating top quark decay. The curves show the SM (black solid), and SM+2HDM (II) with $m_{H^+}=80$ GeV for three different values of $\tan\beta$: 50 (dotted), 25 (short-dashed), and 10 (long-dashed). 
$\mathcal{BR}(H^+\to\tau^+\nu_\tau)=1$ was assumed for all values of
$\tan\beta$. The contours are equivalent in the low $\tan\beta$ regime, whenever $\mathcal{BR}(t\to bH^+)\times\mathcal{BR}(H^+\to \tau^+\nu_\tau)$ is the same.}
\label{Fig:thetastar}

\end{centering}
\end{figure}
In Figure~\ref{Fig:thetastar}, we show the angular distribution of $\tau$ leptons in the rest system of the boson mediating the $t(\bar{t})$ decay. We show here the expectations for the SM, given by Equation~\ref{Eq:ldistW}, and mixtures of SM+2HDM (II) with $\mathcal{BR}(t\to bH^+)$ corresponding to three different values of $\tan\beta$. It is assumed either that $m_{H^+}=80$ GeV, or that kinematic effects can be compensated for on event-by-event basis. Even for $\mathcal{BR}(t\to bH^+)\lesssim 0.1$, the $H^\pm$ events give a significant contribution since $\mathcal{BR}(W^\pm\to \tau^\pm\nu_\tau)=0.1125$ is equally small.

\subsection{Polarization Observables}
From the matrix elements (\ref{Eq:rhoW}) and (\ref{Eq:rhoH}) some useful hints
are obtained on how the spin is analyzed in top quark decays. The spin-dependent
term, proportional to $\sigma_{\lambda\lambda'}$, appears with different
associated momentum directions in the two channels. The particle with this
momentum will analyze the spin most effectively in the corresponding
case. To see how this comes about, recall \cite{Jezabek&Kuhn:NPB:1988} how the
spin of a given top quark is analyzed. The decay products will
experience angular distributions reflecting the spin state of the parent. Each
polarized partial width $\Gamma(t_\uparrow\to bX^+\to bl^+\nu_l)$ of a decaying
fermion can be put in the form
\begin{equation}
\label{Eq:dGamma}
\frac{1}{\Gamma}\frac{\mathrm{d}\Gamma}{\mathrm{d}\cos\theta_i}=\frac{
1+\alpha_i\cos\theta_i}{2},
\end{equation}
where $\theta_i$ denotes the angle of decay product $i$ to the spin helicity
axis, calculated in the rest frame of the decaying particle. The \emph{spin
analyzing coefficients} $\alpha_i$ determine the efficiency of a given particle
to analyze the spin of the parent. The factorization of $\Gamma$ in
energy-dependent and angular parts holds to a high degree also when including
radiative QCD corrections \cite{Jezabek&Kuhn:NPB:1988}. 

To obtain the full set of $\alpha_i$ for a given decay, it is necessary to integrate the polarized
matrix elements. Using the kinematic variables $x=2p\cdot k_2/m_t^2$ and
$y=(k_2+k_3)^2/m_t^2$, the Dalitz parametrization
\cite{Czarnecki&Jezabek:NPB:1994} of the 3-body phase space is written as
\begin{equation}
\label{Eq:PS}
\mathrm{d}\Phi_3=\frac{1}{(2\pi)^5}\mathrm{d}x\mathrm{d}y\mathrm{d}\gamma\mathrm
{d}\beta\mathrm{d}\cos\theta.
\end{equation}
The three Euler angles $\gamma$, $\beta$, and $\theta$ are here chosen according to
\cite{Czarnecki&al:NPB:1991}, so that $\theta$ coincides with the helicity angle
$\theta_i$ discussed above. The integration over $\beta$ is always trivial and
gives $2\pi$. The integration over $\gamma$ is non-trivial only for two,
spin-dependent, quantities. With positive spin projection along the helicity
axis, the result is
\begin{equation*}
\int\mathrm{d}\gamma\ s\cdot k_1=2\pi \cos\theta
\left[\frac{1}{2}\left(1+y-\delta^2\right)-\frac{y}{x}\right]m_t^2\\
\end{equation*}
and since $\sum_i k_i\cdot s=p\cdot s=0$, with $k_2\cdot s = -\cos\theta
\frac{x}{2}m_t^2$, the other interesting integral becomes
\begin{equation*}
\int\mathrm{d}\gamma\ s\cdot k_3=2\pi
\cos\theta\left[\frac{y}{x}-\frac{1}{2}(1+y-x-\delta^2)+\right]m_t^2.
\end{equation*}

The coefficients $\alpha_i$ have been determined for both decay channels. We
summarize our results in Table \ref{tab:alphas} for a decaying $t$ with positive
helicity. Expressions for the other helicity state, or for the charge conjugate
$\bar{t}$ decay, are obtained by an overall change of sign. Our results agree with those
presented in \cite{Mahlon&Parke:PRD:1996}, except that their expressions for the
scalar case do not contain the factor $(A^2-B^2)/(A^2+B^2)f(\xi,A,B)$. This
factor contains all the dependence of the spin analyzing power on the Lorentz
structure of the coupling. We note further that the expression we obtain for $\alpha_H$ is
in agreement with that of \cite{Korner&Mauser:hep-ph:2002}, where also
$\mathcal{O}(\alpha_s)$ corrections to this quantity are given. The inclusion of NLO corrections does not modify the $\tan\beta$ dependence of the $\alpha_i$, even if the numerical values are slightly altered. Since we aim to compare the analytic results to a LO Monte Carlo simulation, we use only the LO results for $\alpha_i$ throughout this work.
 \begin{table}
\centering
 \caption{Spin analyzing coefficients $\alpha_i$ for different decay products in
SM and scalar decay of a polarized top quark ($t_\uparrow\to bW^+/H^+\to
bl^+\nu_l$), or equivalently ($t_\uparrow\to bW^+/H^+\to b\bar{d}u$).}         
\begin{tabular}{ccc}
   \hline       
         Analyzing & \multicolumn{2}{c}{Decay channel} \\
         
          particle & $W^+~(\omega=m_W^2/m_t^2)$ & $H^+~(\xi=m_{H^+}^2/m_t^2)$ \\
         \hline
          & & \\[-3pt]        
         $b$  & $-\dfrac{1-2\omega}{1+2\omega}$ & $-\dfrac{A^2-B^2}{A^2+B^2}f(\xi,A,B)$ \\[8pt]
         $W^+/H^+$  & $\dfrac{1-2\omega}{1+2\omega}$ & $\dfrac{A^2-B^2}{A^2+B^2}f(\xi,A,B)$ \\[8pt]
         $l^+~(\bar{d})$  & $1$ & $\dfrac{1-\xi^2+2\xi\ln\xi}{(1-\xi)^2}\dfrac{A^2-B^2}{A^2+B^2}f(\xi,A,B)$\\[8pt]
         $\nu_l~(u)$  & $\dfrac{(1-\omega)(1-11\omega-2\omega^2)-12\omega^2\ln\omega}{(1-\omega)^2(1+2\omega)}$ 
         & $-\dfrac{1-\xi^2+2\xi\ln\xi}{(1-\xi)^2}\dfrac{A^2-B^2}{A^2+B^2}f(\xi,A,B)$\\[12pt]
         \hline
\end{tabular}
\label{tab:alphas}
\end{table}

Since the top quark spins are not directly observable themselves, what will be
accessible are quantities constructed only from the final state momenta. The
most direct such being the doubly differential distributions of the same type as in Equation~(\ref{Eq:dGamma}), but now involving two particles $(i,j)$; one
from each decaying top quark. The helicity angles $\theta_i$ and $\theta_j$ are
then calculated in the rest systems of the respective parents\footnote{We use
the convention of performing rotation-free boosts from the $\ttbar$ CM system to
define the orientation of the $t$$(\bar{t})$ rest systems. Alternatively, one
could perform a rotation-free boost directly from the hadronic CM system.}. The
most general expression of this type is
\begin{equation}
\label{Eq:ddiff1}
\frac{1}{N}\frac{\mathrm{d}^2N}{\mathrm{d}\cos\theta_i\mathrm{d}\cos\theta_j}
=\frac{1}{4}\Bigl(1+P_1\alpha_i\cos\theta_i+P_2\alpha_j\cos\theta_j+\mathcal{C}
\alpha_i\alpha_j\cos\theta_i\cos\theta_j\Bigr)
\end{equation}
where $P_1(P_2)$ measures the degrees of transverse polarization of the
$t(\bar{t})$. $\mathcal{C}$ is the correlation parameter discussed in
Section~\ref{Sect:Production}. In leading order QCD, with the spin quantized in
the helicity basis, $P_1=P_2=0$ by parity invariance. Instead of
(\ref{Eq:ddiff1}) the simpler distribution
\begin{equation}
\label{Eq:ddiff2}
\frac{1}{N}\frac{\mathrm{d}^2N}{\mathrm{d}\cos\theta_i\,\mathrm{d}\cos\theta_j}
=\frac{1}{4}\Bigl(1+\mathcal{C}\alpha_i\alpha_j\cos\theta_i\cos\theta_j\Bigr)
\end{equation}
is therefore expected to obtain. 

It is also possible to form one-dimensional distributions, \emph{e.g.} as studied in \cite{Bernreuther&al:NPB:2004}. If we define the angle $\theta_{ij}$ between the vectors $i$ and $j$ as in the previous distribution, we get
\begin{equation}
\label{Eq:diffct}
\frac{1}{N}\frac{\mathrm{d}N}{\mathrm{d}\cos\theta_{ij}}=\frac{1}{2}\left(1+\mathcal{D}\alpha_i\alpha_j\cos\theta_{ij}\right).
\end{equation}
Here the coefficient $\mathcal{D}$ is related to $\mathcal{C}$, but in general it has a different value. To determine $\mathcal{C}$ and $\mathcal{D}$ from angular distributions, the relations
\begin{eqnarray} 
\mathcal{C}&=&\frac{9}{\alpha_i\alpha_j}\Ave{\cos\theta_i\cos\theta_j}\label{detC}\\
\mathcal{D}&=&\frac{3}{\alpha_i\alpha_j}\Ave{\cos\theta_{ij}}\label{detD}
\end{eqnarray}
can be used. Conversely, when the correlation coefficients are known, these relations can be used to determine the product $\alpha_i\alpha_j$. With a leading order $\mathcal{C}=0.319$, we obtain the corresponding $\mathcal{D}=-0.216$.

\subsection{Analysis in 2HDM (II)}
\label{Sect:2HDMnum}
\begin{figure}
\begin{centering}
\includegraphics[width=0.5\columnwidth, keepaspectratio]{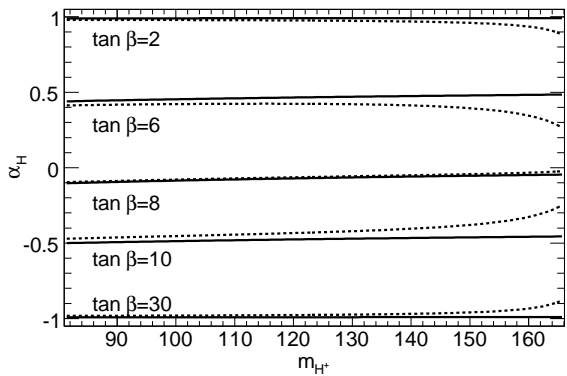}
\caption{Influence of the threshold factor on the spin analyzing coefficients. The dashed curves show the full expression for $\alpha_H$ while
the solid curves show $\alpha_H$ in the approximation $f(\xi,A,B)=1$}
\label{Fig:thold}
\end{centering}
\end{figure}

As an illustrative example of the differences between the SM and a new scalar
decay, let us consider in some detail the results for a 2HDM (II) with $H^\pm$
couplings from Table~\ref{tab:ABC}. In all the following, we shall fix the top
mass to $m_t=170.9$ GeV \cite{Topmass:2007}. Starting by analyzing the threshold
region $m_{H^+}\simeq m_t-m_b$, Figure~\ref{Fig:thold} shows $\alpha_H$ as a
function of $m_{H^+}$ for different values of $\tan\beta$. We see that the
threshold suppression becomes significant as $m_{H^+}$ approaches the kinematic
limit. However, for very large, very small or intermediate ($\sim 7$)
$\tan\beta$ values, we infer the threshold correction to be less than $10\%$
also for $m_{H^+}=160$ GeV. In these regions of parameter space, the threshold
factor can be effectively ignored. Note that this argument is not specific to $\alpha_H$, but applies to all $\alpha_i$, since $f(\xi,A,B)$ is a universal factor. 

\begin{figure}
\begin{centering}

\subfigure{
   \label{Fig:alphaA}
   \includegraphics[width=0.45\columnwidth,keepaspectratio]{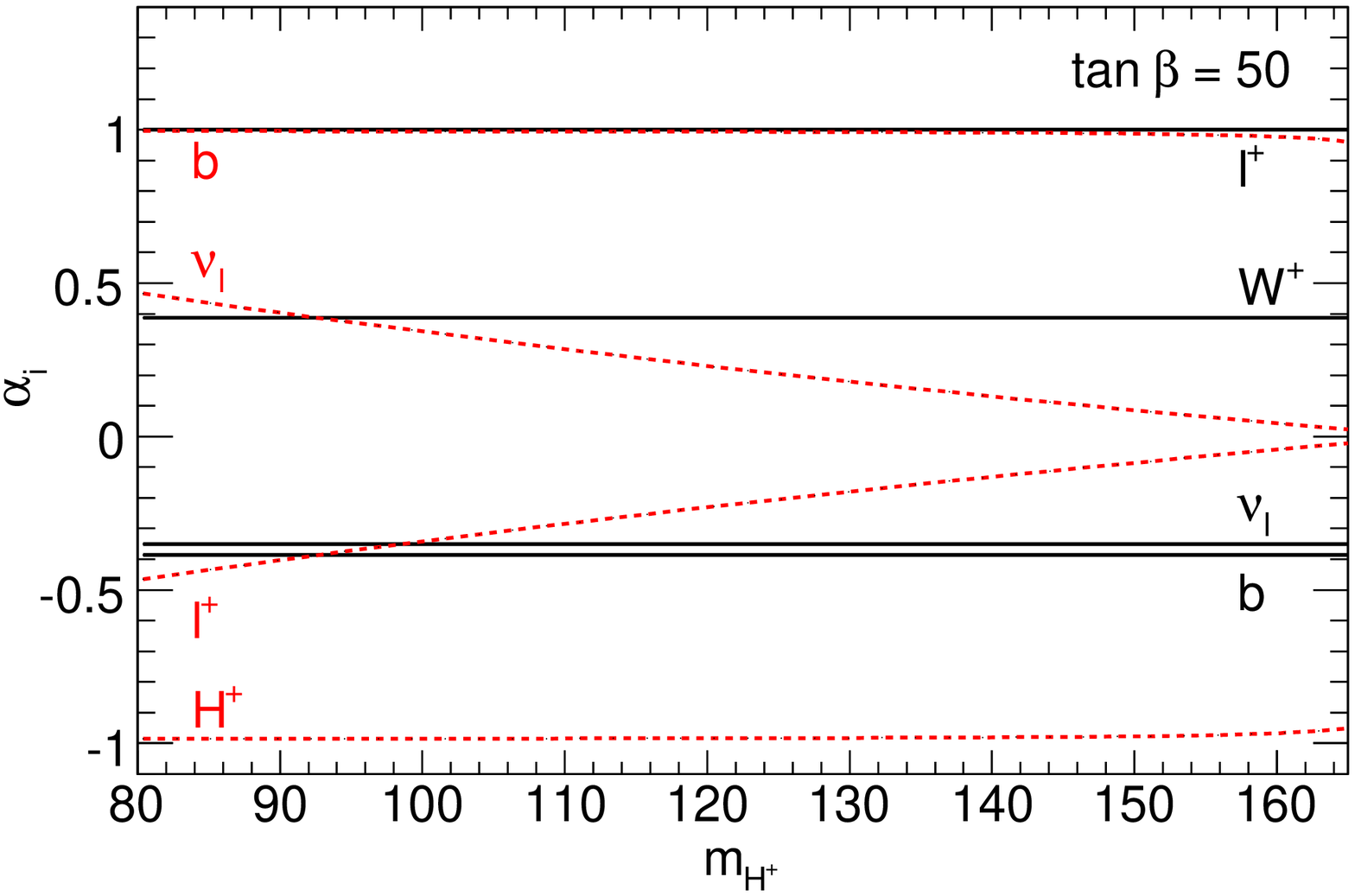}

}
\subfigure{
   \label{Fig:alphaB}
   \includegraphics[width=0.45\columnwidth,keepaspectratio]{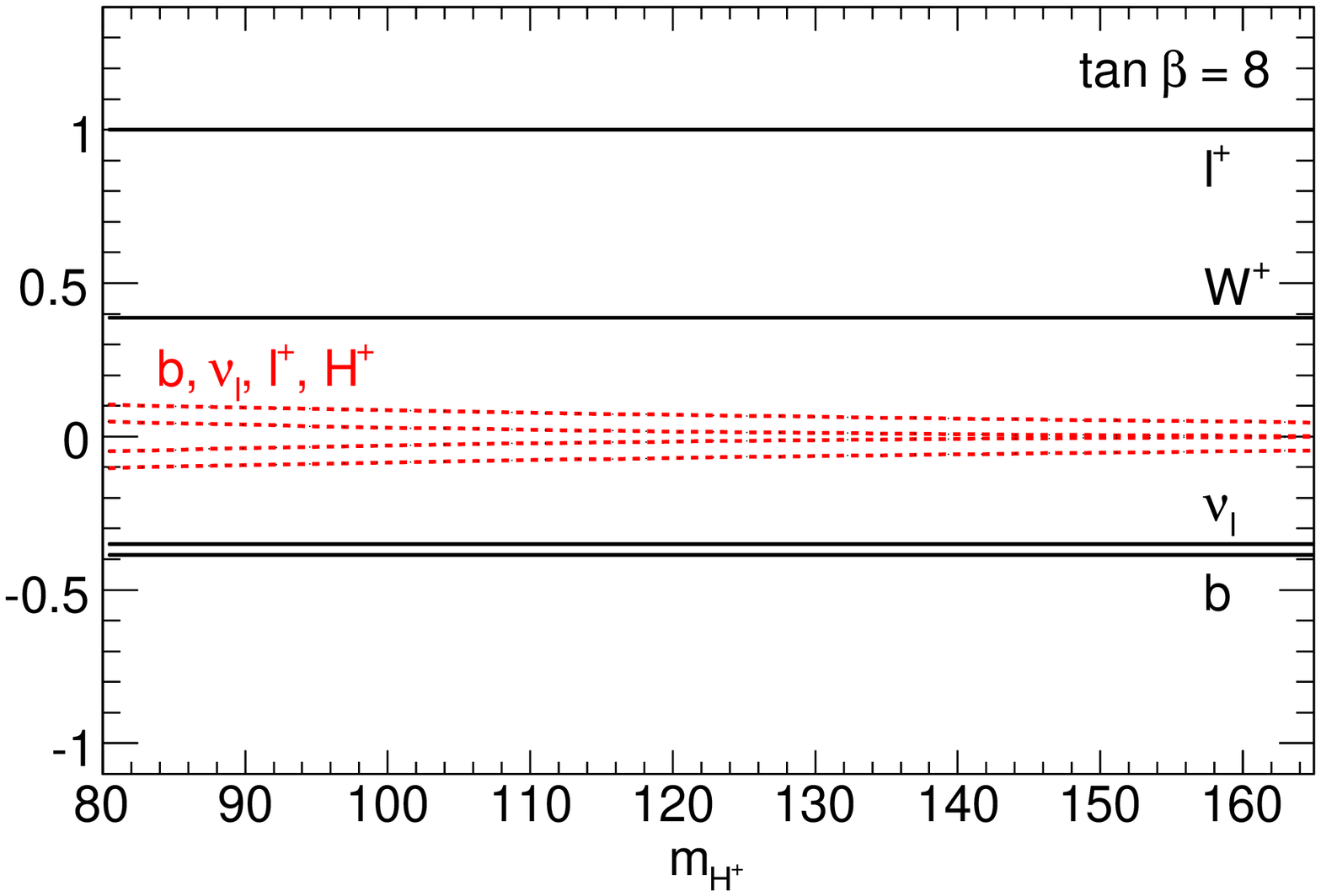}
}
\caption{Spin analyzing coefficients $\alpha_i$ for top decay within SM (solid black), and in the 2HDM (II) (dashed red), presented for $\tan\beta=50$ (left) and $\tan\beta=8$ (right). For low $\tan\beta\sim 1$, the Higgs results presented in the left plot are valid with an overall change of sign.}
   \label{Fig:alpha}
\end{centering}
\end{figure}

Summarizing the results presented in Table~\ref{tab:alphas} for the 2HDM (II),
Figure~\ref{Fig:alpha} shows a numerical evaluation of the analytic expressions for all $\alpha_i$. The results are presented for two values of $\tan\beta$: one large value ($\tan\beta= 50$), for which the
efficiency in analyzing the spin is optimum, and one intermediate value
$\tan\beta=\sqrt{m_t/\overline{m}_b}\sim 8$ (for $\overline{m}_b=3.2$
GeV with $m_{H^+}=100$ GeV), where all the sensitivity to analyze the top spin
vanishes. This value corresponds to a purely scalar coupling, thus to an
isotropic decay. For $\tan\beta\lesssim 8$, all $\alpha_i$ acquire an
extra minus sign compared to $\tan\beta\gtrsim8$. This corresponds to a
shift from predominantly right-chiral to left-chiral coupling.
We see also, that the efficiency to analyze
the top spin is not highest using the charged lepton, as is shown to be the case
for the SM. Instead the most efficient probe is either the Higgs momentum itself
or the associated $b$ quark. This is easily understood; since the $H^\pm$ itself
does not carry any spin, the top spin information can only be transferred to the angular distributions of the $b/H^+$. In a vector decay, parts of this
information go into the different polarization states of the $W$, as discussed above.

\begin{figure}
\begin{centering}
\includegraphics[width=0.5\columnwidth, keepaspectratio]{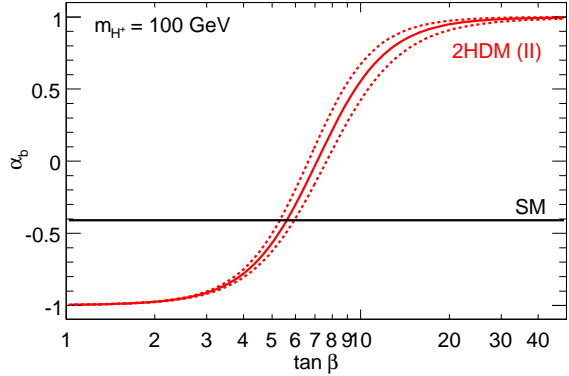}
\caption{The solid red curve shows $\alpha_b$ for the 2HDM (II), whereas the
dashed curves correspond to the SUSY-corrected model ($\twobar$) with
$\epsilon_b=-\epsilon_t'=\pm0.01$. The mass $m_{H^+}=100$ GeV was used. For comparison, the black line shows $\alpha_b$ for the SM decay.}
\label{Fig:A2B2}
\end{centering}
\end{figure}

Figure~\ref{Fig:A2B2} displays $\alpha_b$ as a function of $\tan\beta$ for a
fixed $m_{H^+}=100$ GeV. It illustrates clearly the transition from $\alpha_b=-1$ analyzing
power at small $\tan\beta$, to $\alpha_b=1$ associated with
a right handed coupling for large $\tan\beta$.
Note also, that for $\tan\beta=\sqrt{\sqrt{2}\frac{m_W}{\overline{m}_b}}\sim 6$, the
$b$ quark coefficient $\alpha_b$ of the Higgs events mimics that of the SM
decay. The only dependence in this relation on $m_{H^+}$ enters through the
running mass $\overline{m}_b$. No corresponding value exists where the $\alpha_l$ are equal,
since the Higgs value is bounded by kinematics to $|\alpha_l|\lesssim 0.5$,
whereas the SM value is always $\alpha_l=1$. 

To compare with the case when $\tan\beta$-enhanced SUSY corrections to the
charged Higgs couplings are included, we give in Figure~\ref{Fig:A2B2} also the
values of $\alpha_b$ for the 2HDM ($\twobar$). Rather than to calculate the
corrections for a specific SUSY model spectrum, we parametrize them in terms of
the parameters $|\epsilon_b|\leq 0.01$ and $|\epsilon_t'|\leq 0.01$. These are
reasonable maximum values \cite{Degrassi:JHEP:2000}, which correspond roughly to $\alpha_s/3\pi$ as discussed
above. Figure~\ref{Fig:A2B2} clearly shows that even though these corrections are enhanced
by $\tan\beta$ in the couplings, they have only a small effect on the ratio that
enters the spin analyzing coefficients. In fact, the largest correction is
obtained not in the high $\tan\beta$ limit, but in the transition region around $\tan\beta=8$~--~$20$. Given the observed
smallness of the SUSY effects on the spin analyzing coefficients, it is acceptable to apply the results
from the 2HDM (II) without modification, both in the high and in the low
$\tan\beta$ regimes. However, total rates for $t\to bH^+$ are of course affected
by the differences.

A few words are to be said also about the $\mathcal{O}(\alpha_\mathrm{s})$ corrections calculated in \cite{Korner&Mauser:hep-ph:2002}. Inclusion of these effects leads to modifications of the spin analyzing coefficients in a fashion very similar to the $\tan\beta$ enhanced SUSY corrections discussed above. These corrections are also largest in the intermediate $\tan\beta$ region, where they can reach $20\%$ in magnitude. In the large and small $\tan\beta$ limits, the NLO corrections have negligible impact. Since our results are most interesting in these limits, we will show plots for $\tan\beta=50$ and $\tan\beta=1$. Interpolation to the intermediate range should then be performed with care, remembering the higher order corrections.

\begin{figure}
\begin{centering}
\subfigure{
   \label{Fig:A}
   \includegraphics[width=0.3\columnwidth,keepaspectratio]{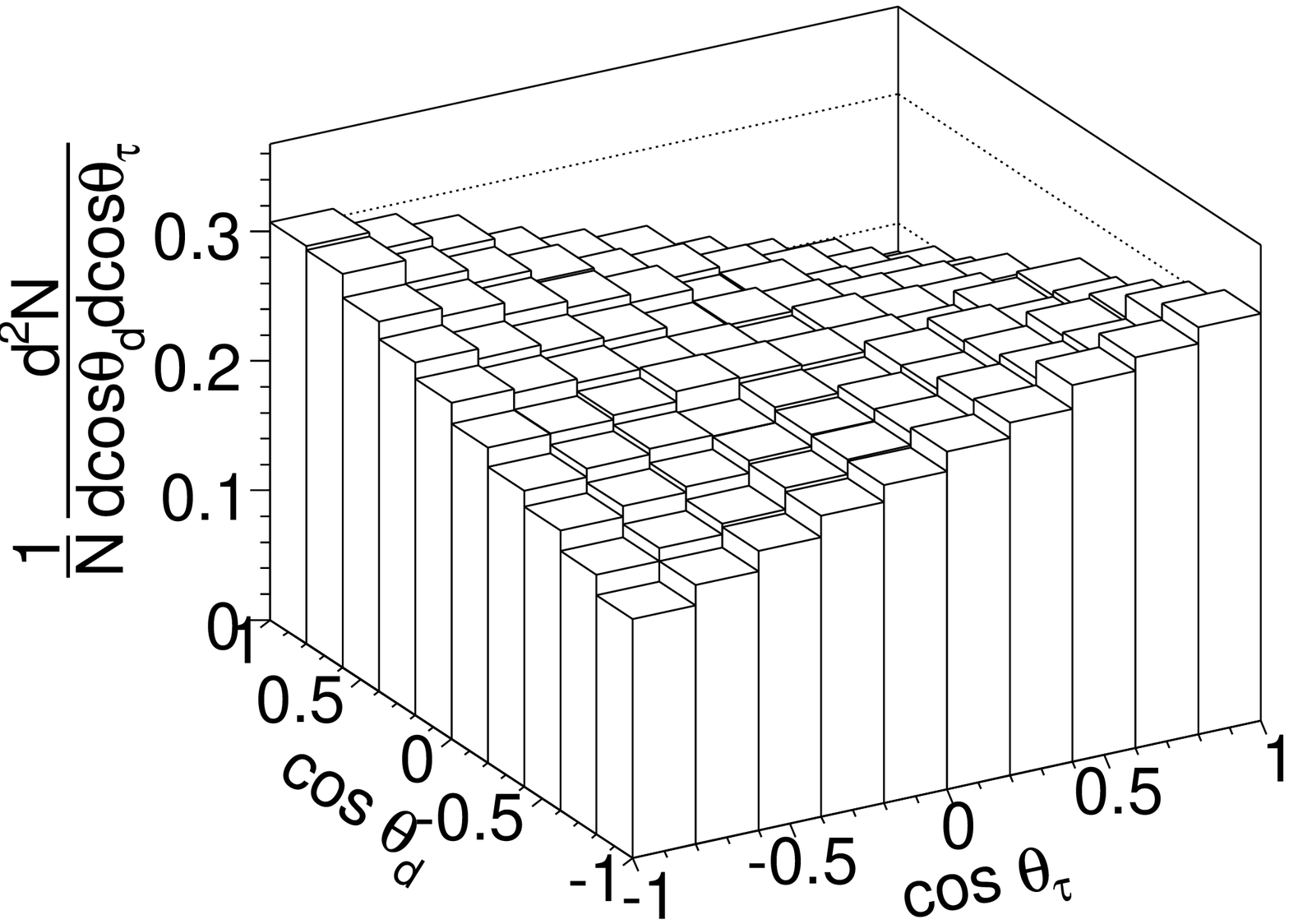}

}
\subfigure{
   \label{Fig:B}
   \includegraphics[width=0.3\columnwidth,keepaspectratio]{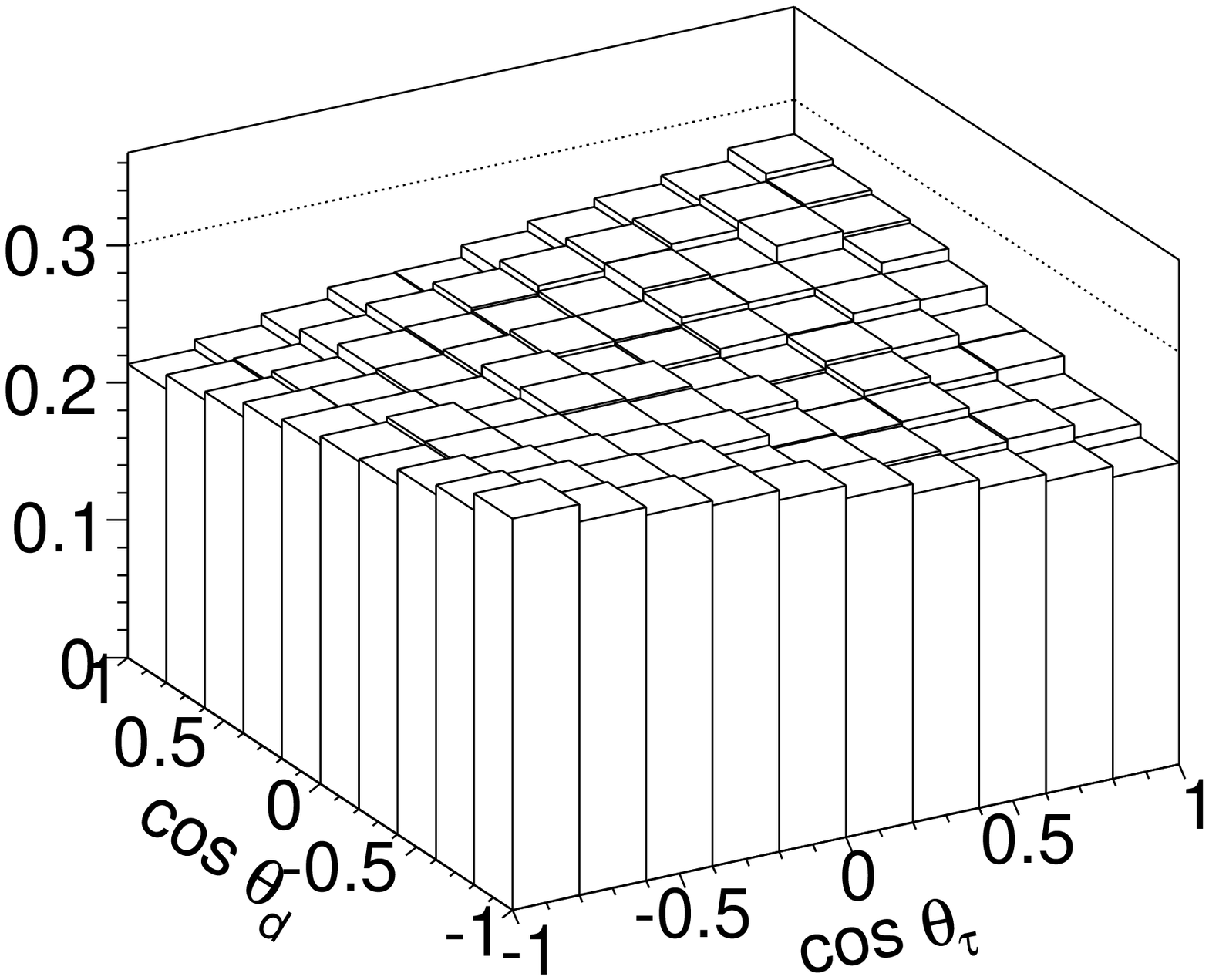}
}
\subfigure{
   \label{Fig:B}
   \includegraphics[width=0.3\columnwidth,keepaspectratio]{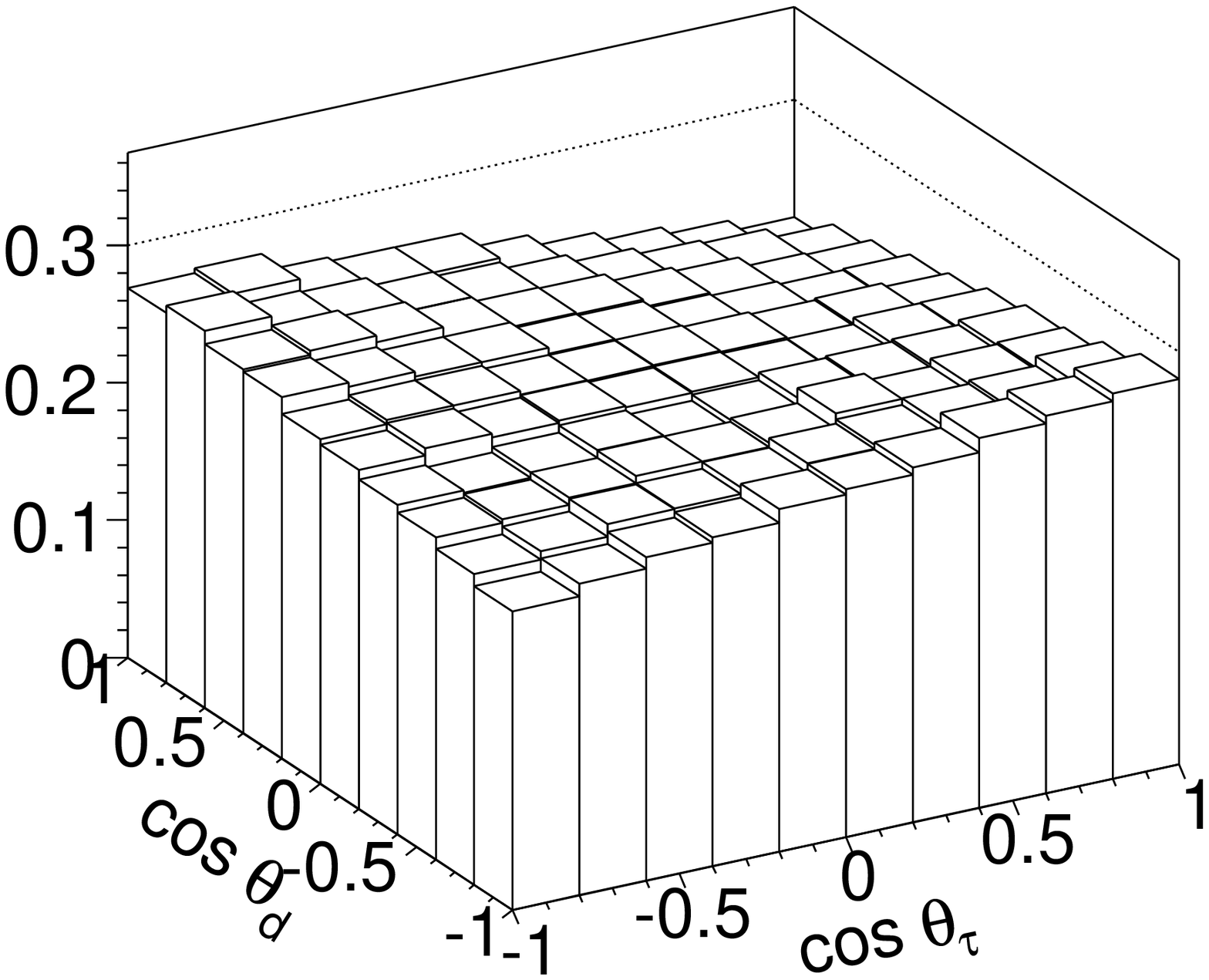}
}
\subfigure{
   \label{Fig:C}
   \includegraphics[width=0.3\columnwidth,keepaspectratio]{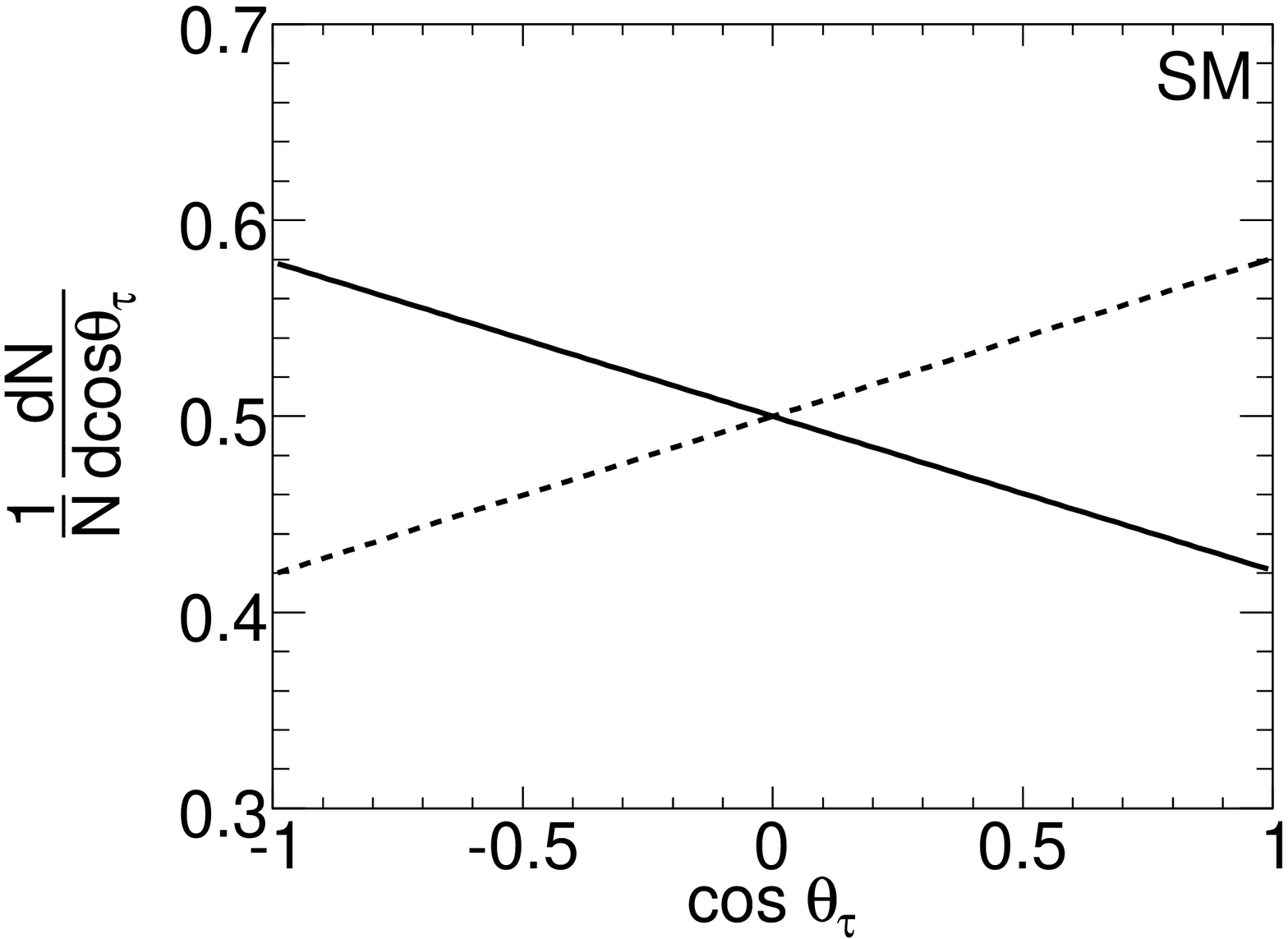}

}
\subfigure{
   \label{Fig:B}
   \includegraphics[width=0.3\columnwidth,keepaspectratio]{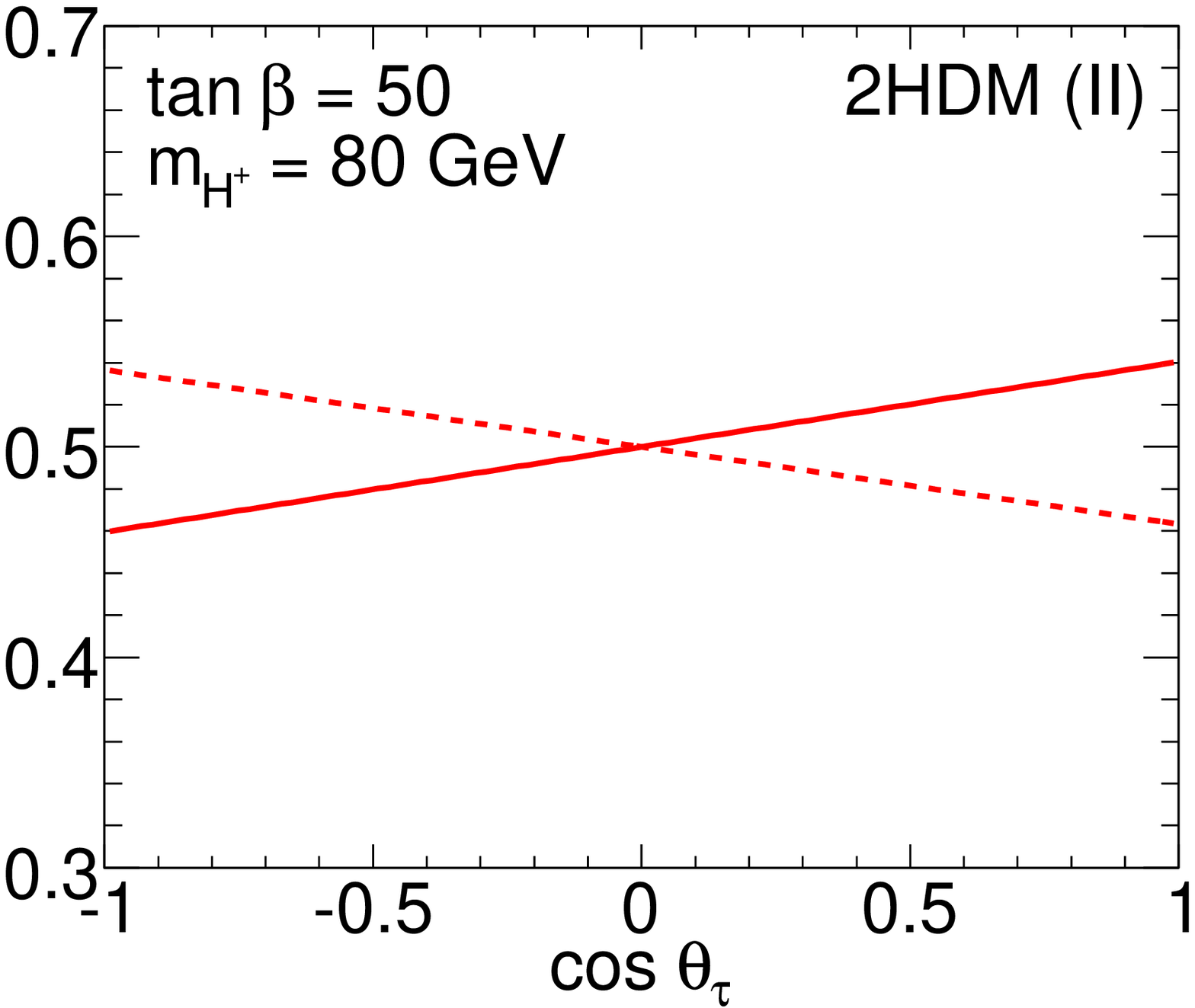}
}
\subfigure{
   \label{Fig:B}
   \includegraphics[width=0.3\columnwidth,keepaspectratio]{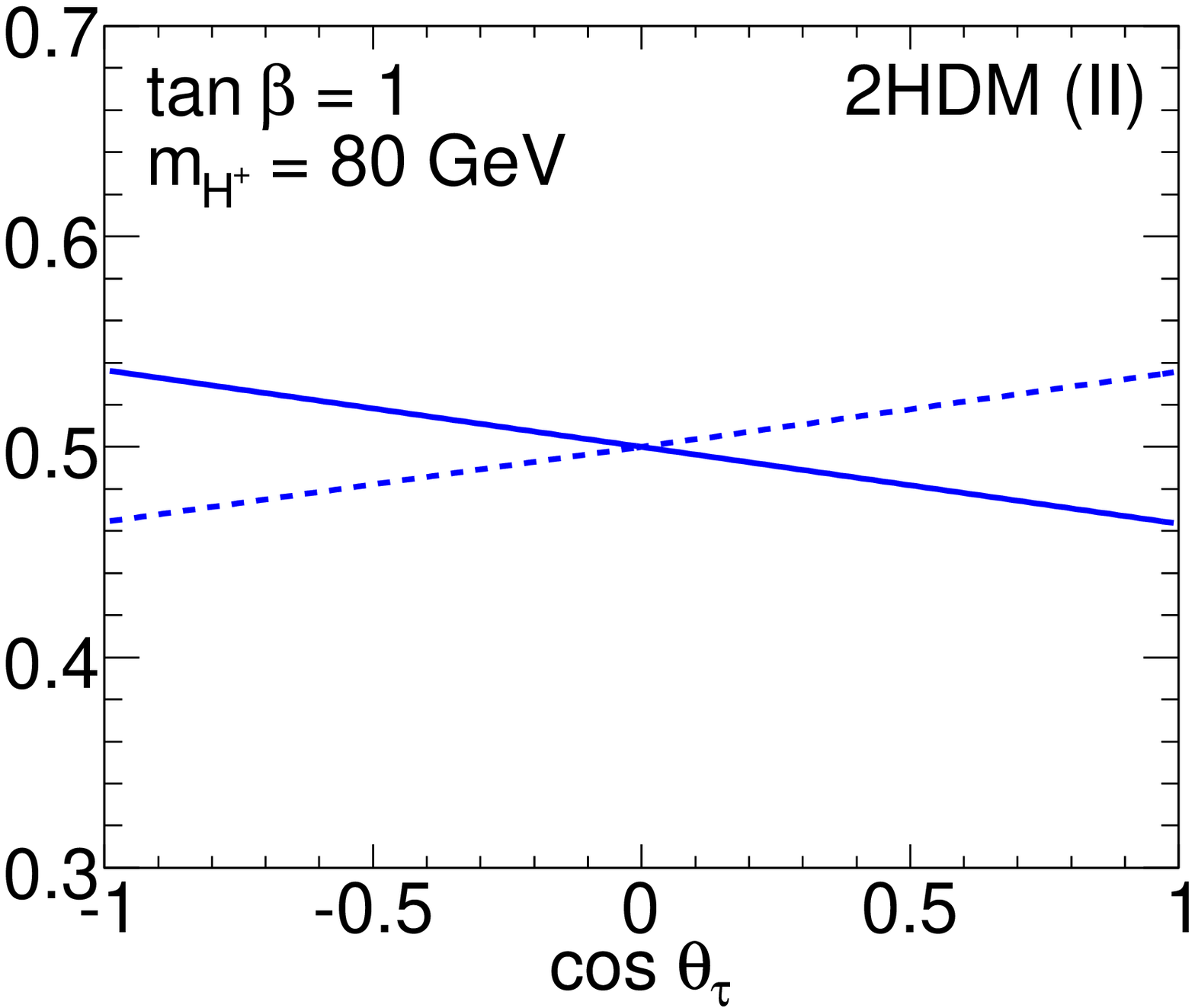}
}
\caption{Upper row shows doubly differential distributions in $\cos\theta_{i,j}$
(defined in the text) of the decay products $d$ and $\tau$ from $\ttbar$
decaying through $W^\pm W^\mp$ (left) and $H^\pm W^\mp$ (center, right). Results
obtained at ME level for $m_{H^+}=80$ GeV, $\tan\beta=50$ (center) and
$\tan\beta=1$ (right). The bottom panel lines correspond to projected angular
distributions with $\cos\theta_d>0$ (solid) and $\cos\theta_d<0$
(dashed).}
   \label{Fig:ll}
\end{centering}
\end{figure}
As a final result for the 2HDM (II) on matrix element level, we show the 
differential distributions in $(\cos\theta_{i},\cos\theta_j)$ and $\cos\theta_{ij}$, described by
Equations~(\ref{Eq:ddiff2}) and (\ref{Eq:diffct}) respectively. The top row in Figure~\ref{Fig:ll} shows
lepton-lepton correlations, where the lepton ($d$ quark) from a $W$ decay is correlated with another lepton (in this case a $\tau$) from $W^\pm$ or $H^\pm$ decay from the opposite side of the event. In the absence of spin correlations, this distribution is expected to be flat. From left to right, Figure~\ref{Fig:ll} gives the results for the SM, for the 2HDM (II) with $\tan\beta=50$, and similarly with $\tan\beta=1$. In the bottom row,
distributions in one of the angles are given. These are obtained using the
other angle to determine the parent spin by applying the
projections $\cos\theta_d>0$ and $\cos\theta_d<0$ to the 2D distributions. Figure~\ref{Fig:ll}
illustrates the points we have previously made about the 2HDM (II), namely that
 i) the lepton-lepton correlation is more efficient in the SM since
$\alpha_l\neq 1$ in the Higgs case, ii) there is a change of sign in
going from high to low values of $\tan\beta$, and iii) the low $\tan\beta$
case is more SM-like.

\begin{figure}
\begin{centering}
\subfigure{
   \label{Fig:A}
   \includegraphics[width=0.3\columnwidth,keepaspectratio]{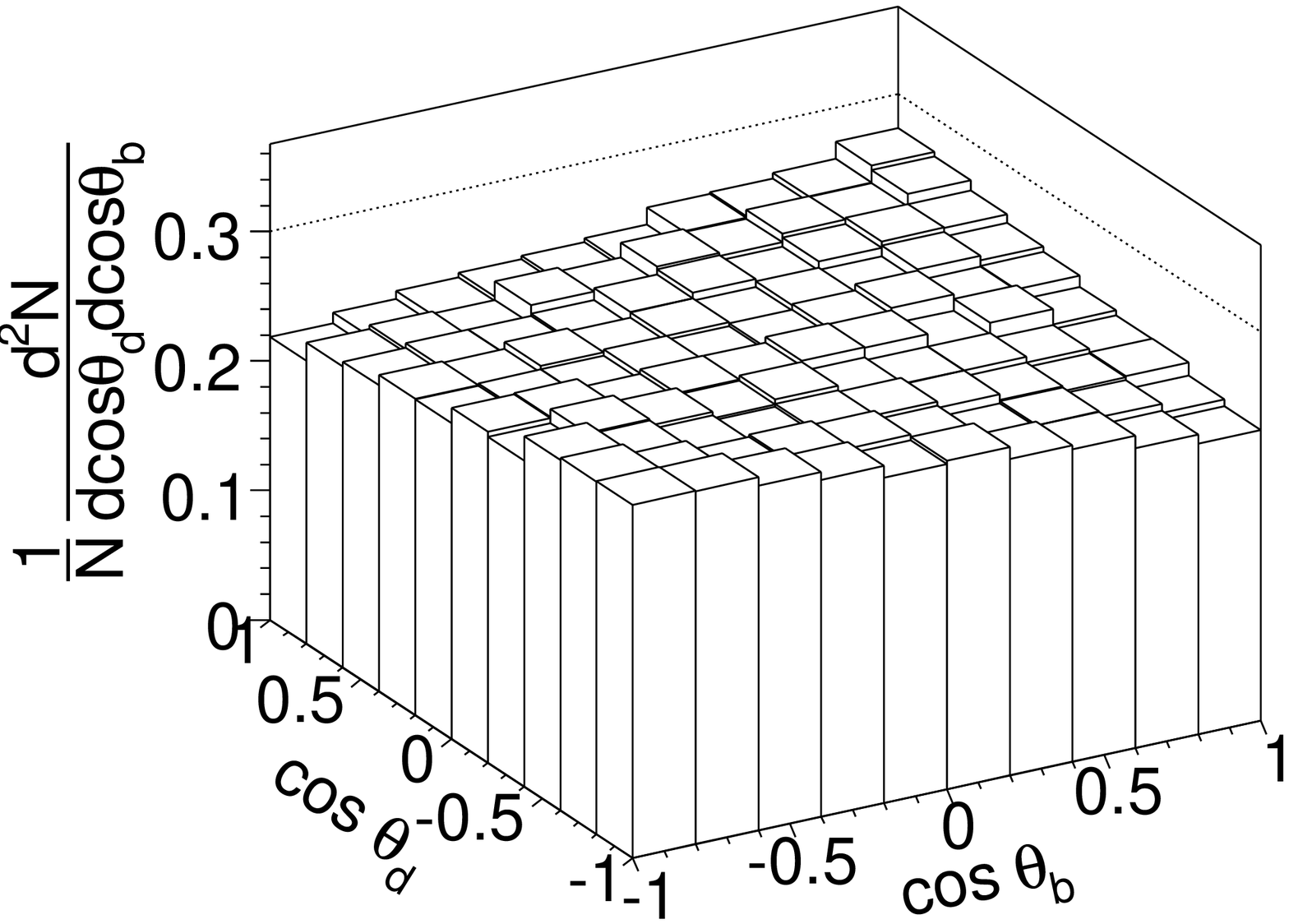}

}
\subfigure{
   \label{Fig:B}
   \includegraphics[width=0.3\columnwidth,keepaspectratio]{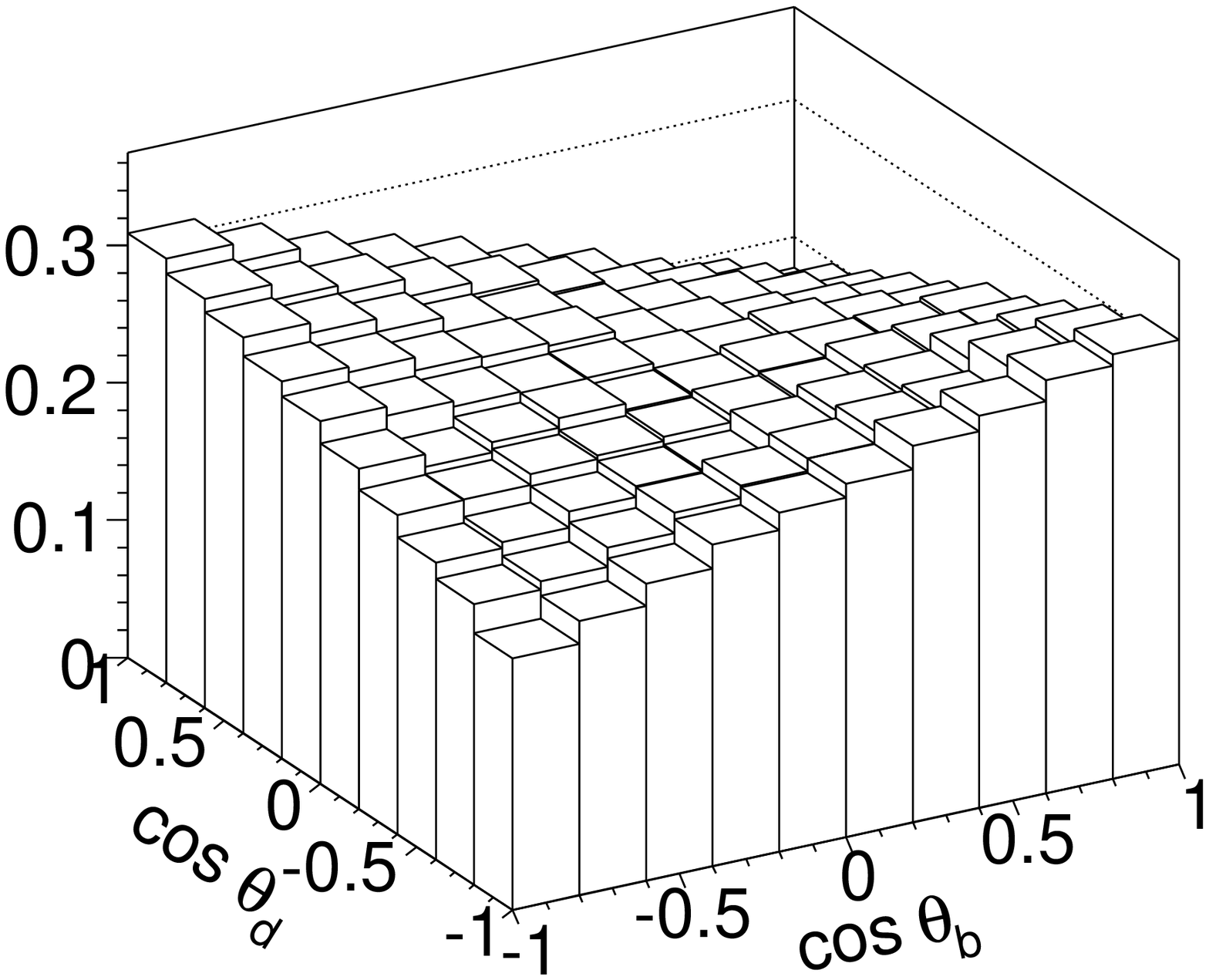}
}
\subfigure{
   \label{Fig:B}
   \includegraphics[width=0.3\columnwidth,keepaspectratio]{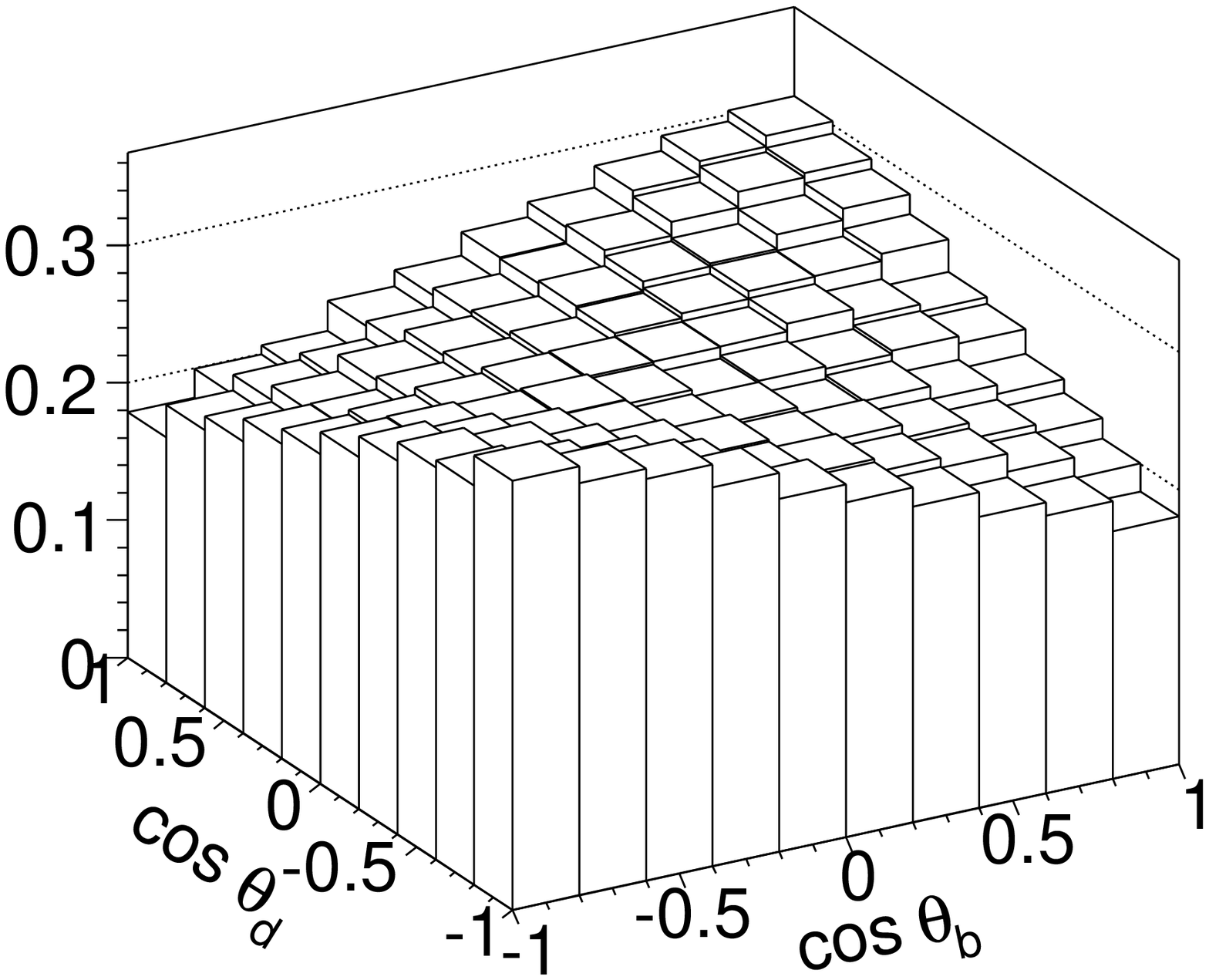}
}
\subfigure{
   \label{Fig:C}
   \includegraphics[width=0.3\columnwidth,keepaspectratio]{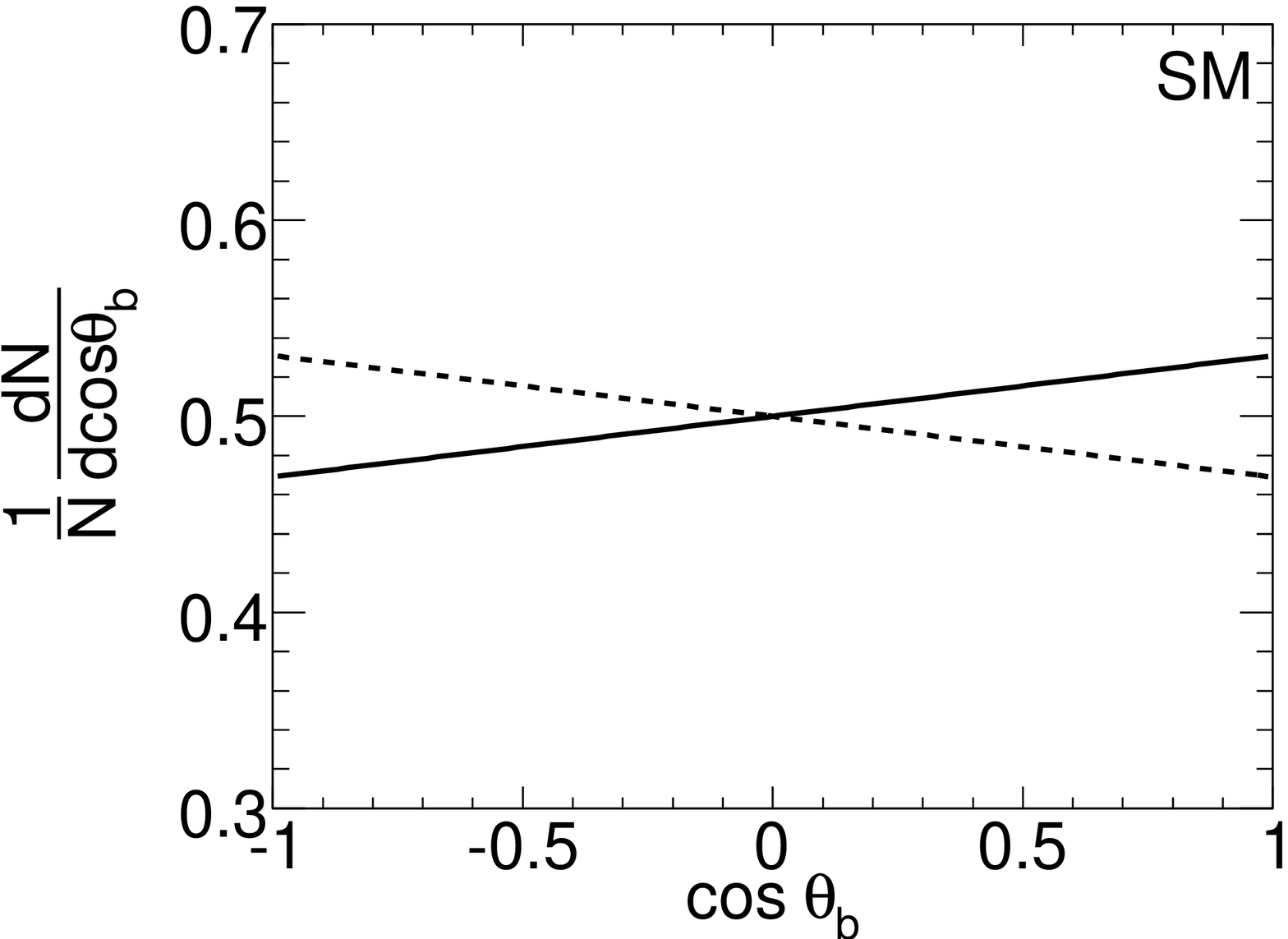}

}
\subfigure{
   \label{Fig:B}
   \includegraphics[width=0.3\columnwidth,keepaspectratio]{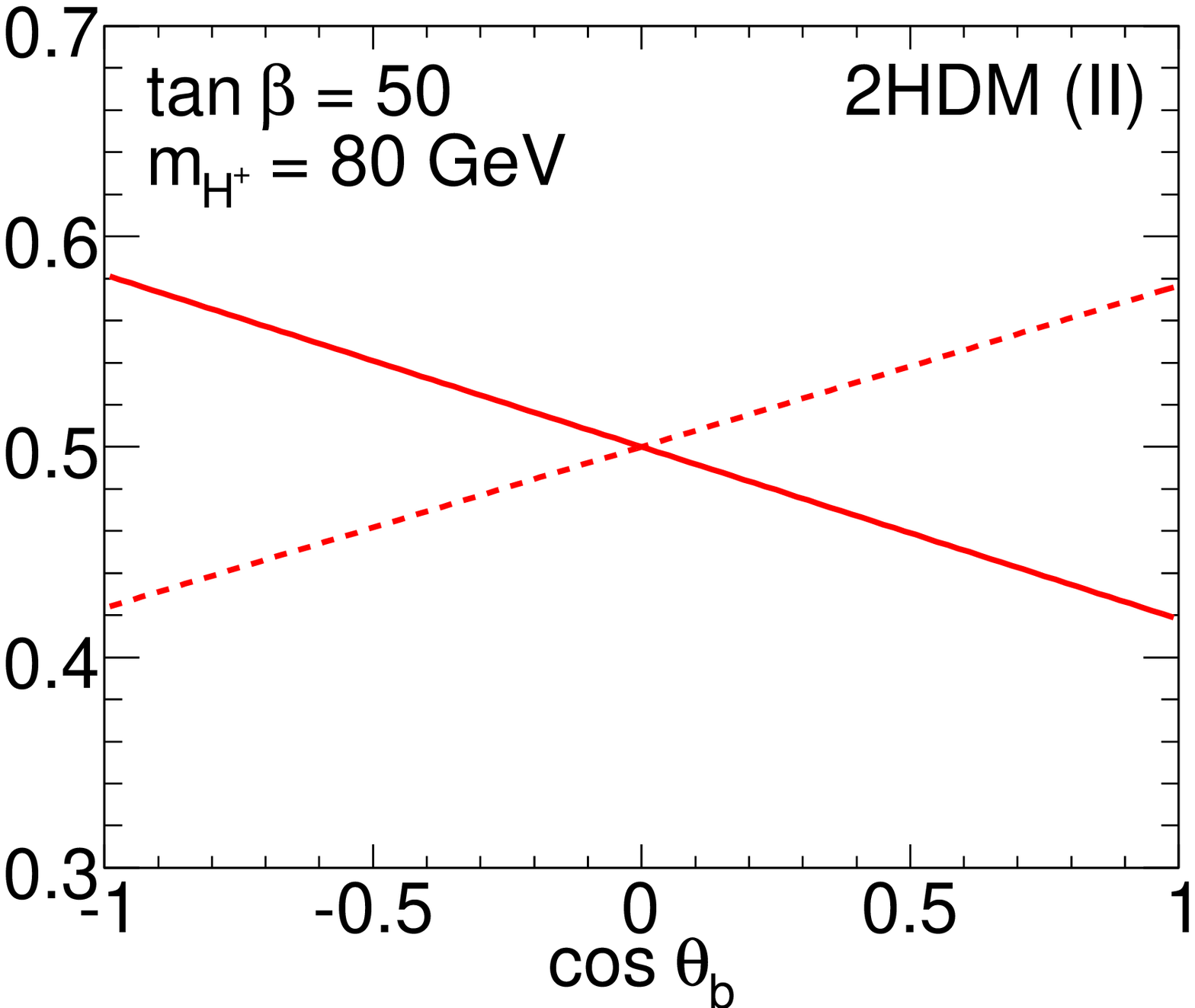}
}
\subfigure{
   \label{Fig:B}
   \includegraphics[width=0.3\columnwidth,]{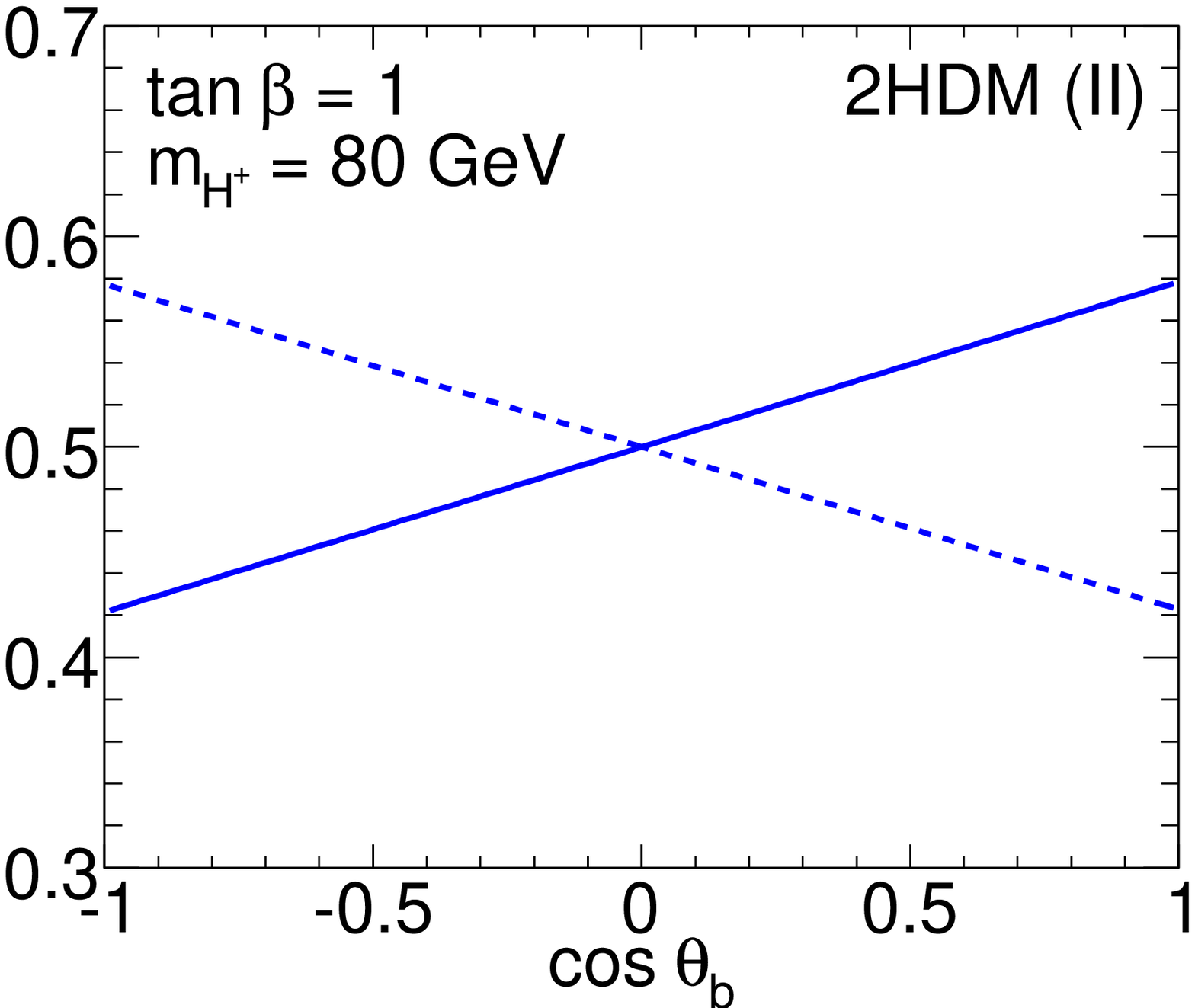}
}
\caption{Angular distributions defined similarly to those in Figure~\protect \ref{Fig:ll}, with the
$\tau$ replaced by the associated $b$ quark originating from the same
$t$($\bar{t}$) decay. Again $m_{H^+}=80$ GeV and the ordering is
SM (left), $\tan\beta=50$ (center), and $\tan\beta=1$ (right).}
   \label{Fig:bl}
\end{centering}
\end{figure}

Going to Figure~\ref{Fig:bl}, the $\tau$ lepton has been
replaced with the $b$ quark associated with the same (Higgs) side of the event.
The three distributions are defined similarly to those in Figure~\ref{Fig:ll}, as are
the projections. From Figure~\ref{Fig:bl}, the increased spin analyzing efficiency when using the associated $b$ quark in the Higgs case is evident.

\begin{figure}
\begin{centering}
\subfigure{
   \label{Fig:A}
   \includegraphics[width=0.45\columnwidth,keepaspectratio]{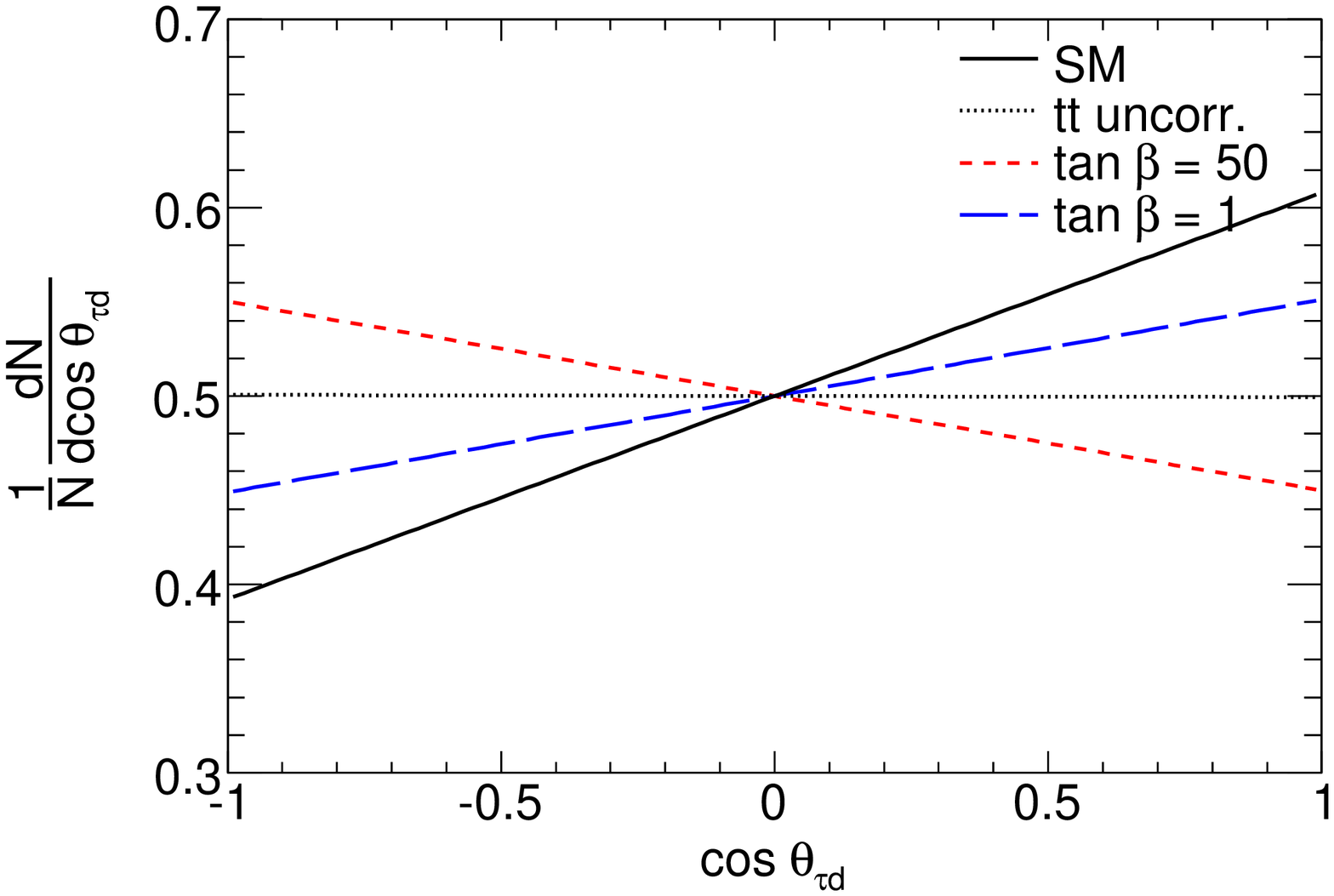}

}
\subfigure{
   \label{Fig:B}
   \includegraphics[width=0.45\columnwidth,keepaspectratio]{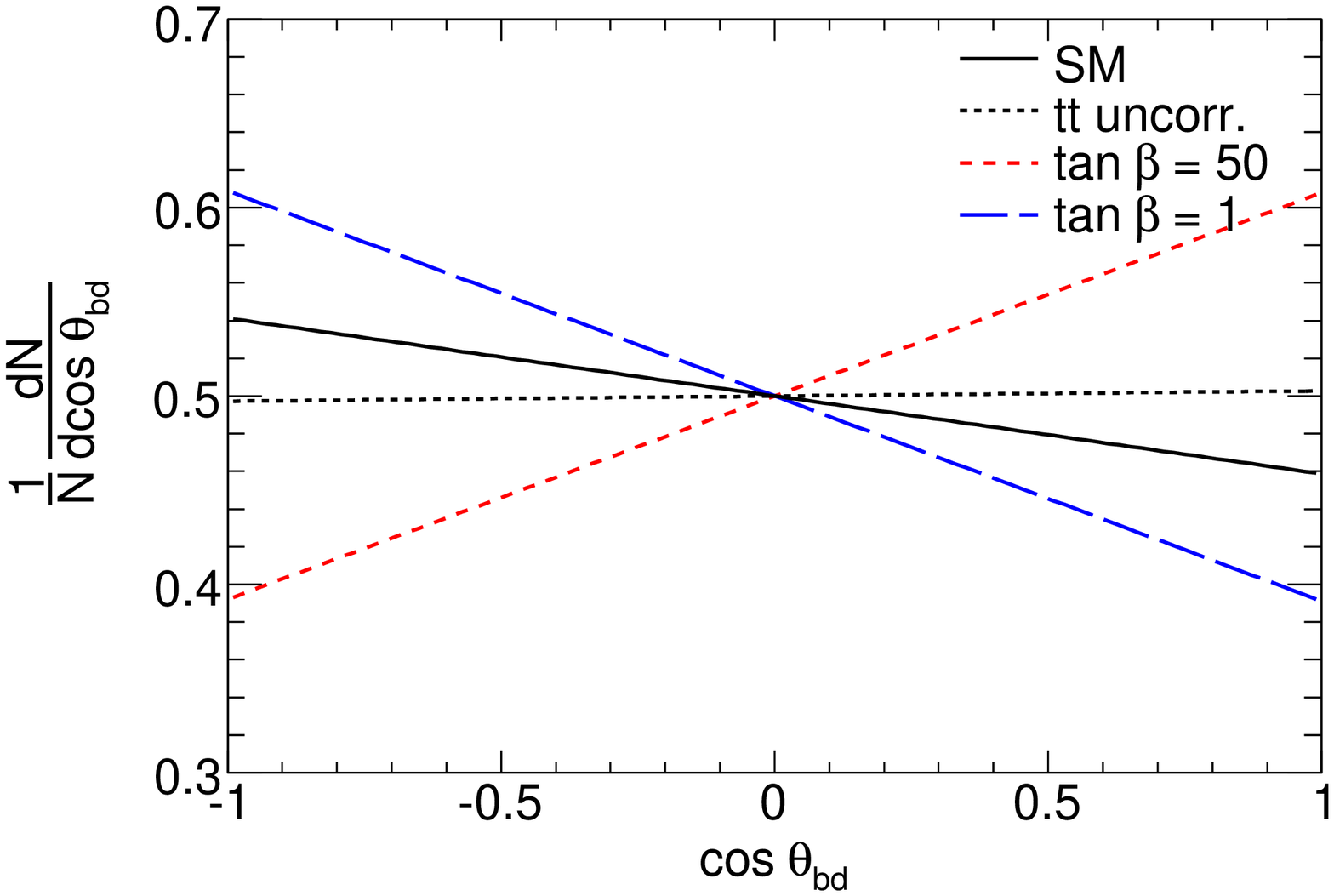}
}
\subfigure{
   \label{Fig:A}
   \includegraphics[width=0.45\columnwidth,keepaspectratio]{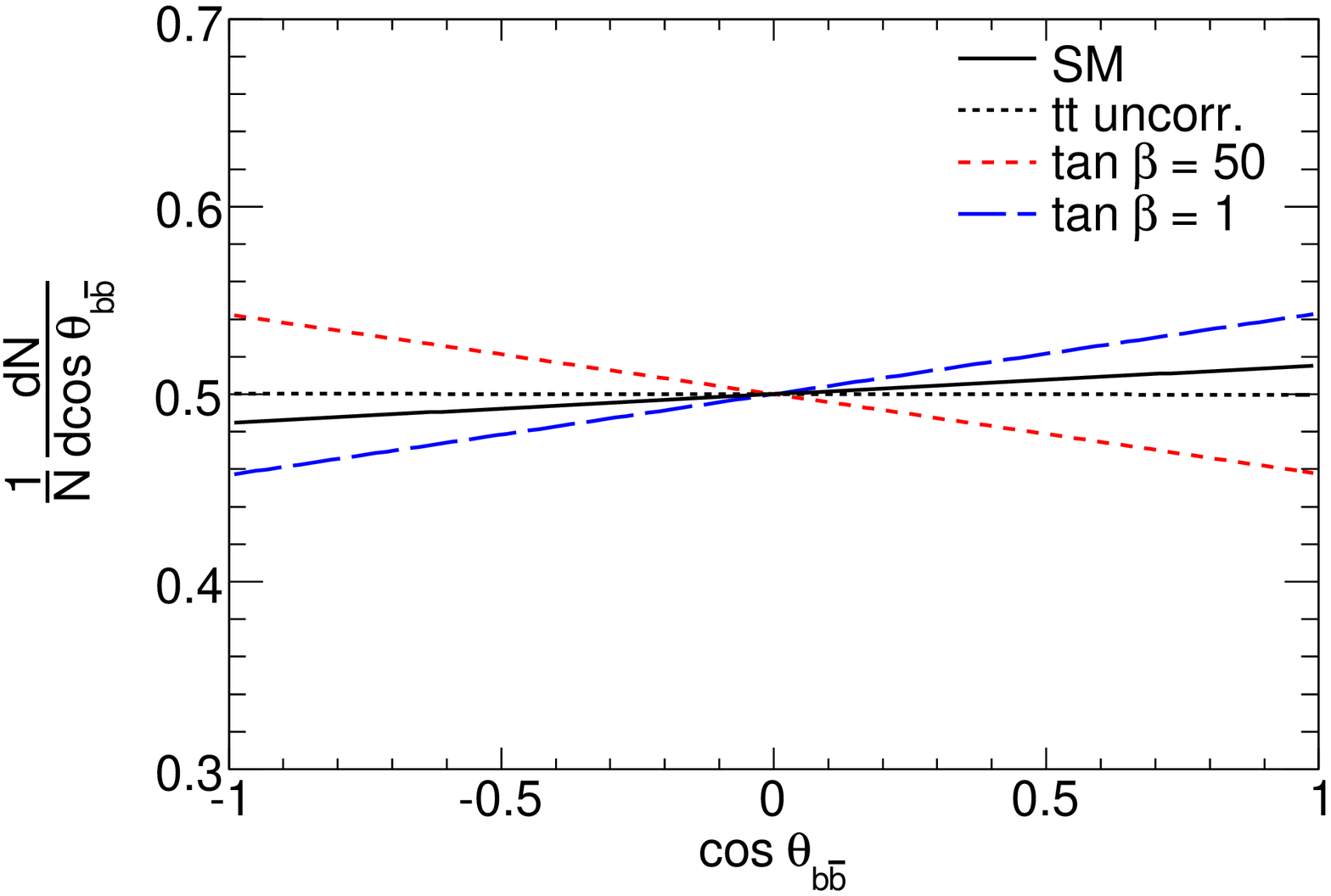}

}
\subfigure{
   \label{Fig:B}
   \includegraphics[width=0.45\columnwidth,keepaspectratio]{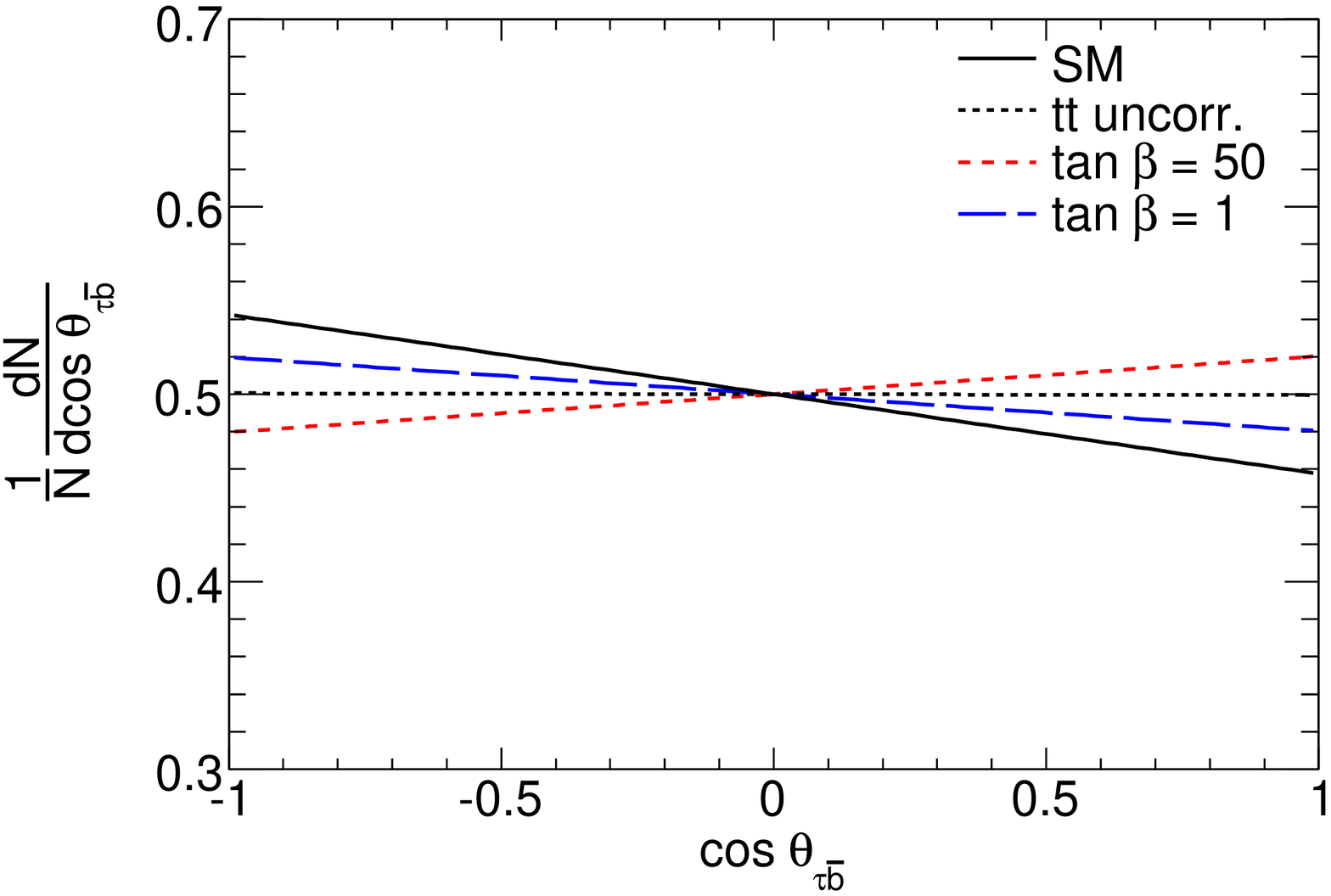}
}
\caption{Distributions at matrix element level of $\cos \theta_{ij}$ for different final state particles $(i,j)$. The dotted line is the expected SM result \emph{without} $\ttbar$ spin correlations. The SM result is shown in black, while the 2HDM (II) is presented for $\tan\beta=1$ (long-dashed blue), and $\tan\beta=50$ (short-dashed red). All results are given for $m_{H^+}=80$ GeV.}
\label{Fig:cost}
\end{centering}
\end{figure}

The last figure to discuss here is Figure~\ref{Fig:cost}, which shows distributions in $\cos\theta_{ij}$ between different particles in the final state. An uncorrelated sample corresponds to a flat line distribution. The specific combinations chosen are $(\tau,d)$, $(b,d)$, $(b,\bar{b})$, and $(\tau,\bar{b})$, where the first particle originates from $H^\pm/W^\pm$ decay, while the second always comes from the opposite side $W^\pm$. The distributions $(H^+,\bar{b})$ and $(H^+,d)$ are equivalent to the distributions $(b,\bar{b})$ and $(b,d)$ displayed in Figure~\ref{Fig:cost} with a change of sign in the spin analyzing coefficient ($\alpha_{H}=-\alpha_b$ in Eq.~(\ref{Eq:diffct})). Hence they would not contribute any additional information. 

By combining in the same plots for the SM with results from the 2HDM, we see directly which correlations are more efficient in the two cases. In agreement with previous results, this is again the $\tau d$ distribution for the SM, and the $b\bar{b}$ distribution for the 2HDM (II), illustrating the universal dependence of the spin correlation effects on $\alpha_i\alpha_j$. We have also found, that when cuts are applied at the parton level, the distributions shown in Figure~\ref{Fig:cost} are not affected nearly as much as those presented in Figures~\ref{Fig:ll} and \ref{Fig:bl}. Using the $\cos\theta_{ij}$ variables to study $\ttbar$ spin correlations is therefore advantageous to avoid the problems with cuts discussed in \cite{Mahlon&Parke:PRD:1996}. We will return to this discussion below.
\section{Monte Carlo Simulations}
\label{Sect:Simulations}
Up to this point, the results we have presented were obtained directly from matrix elements. To give a more realistic assessment of the prospects to observe any of
these spin effects in a collider experiment, complete hadron-level events must be considered. We do this using a Monte Carlo (MC) approach. The framework is again the 2HDM (II) because of its special status as the minimal Higgs model compatible with supersymmetry.

The production of $\ttbar$ is treated completely within the SM. On the decay end, we study 
in parallel the situations when either both decays occur within the SM, or when one of the two top quarks decays through $W^\pm$ and the other through $H^\pm$.\footnote{Of course, there is also the possibility of having both tops in an event decay through $H^\pm$. However, since this mode gives a different final state, and since the effect is sub-leading, we will neglect this contribution here.} Being an interesting process in its own right, the SM decay of $\ttbar$ is also the main irreducible background in searches for light $H^\pm$ \cite{Biscarat&Dosil:ATLAS:2003}. Over nearly the full range of $\tan\beta$, $H^\pm$ decays preferentially to the heaviest lepton available, that is
$H^+\to\tau^+\nu_\tau$. We therefore restrict ourselves to this decay
channel. Consequently, for the SM events, we demand one of the
two $W$ bosons to decay through this mode which has $\mathcal{BR}(W^\pm\to\tau^\pm
\nu_\tau)=0.1125$. The $\tau$ lepton subsequently decays hadronically producing
a $\tau$ jet. For the other $W$, which is always present in the event, we consider the hadronic decay to allow for hadronic top reconstruction.
We show SM and 2HDM results separately, keeping in mind that $\mathcal{BR}(t\to bH^+,H^+\to\tau^+\nu_\tau)$ and $\mathcal{BR}(t\to bW^+,W^+\to\tau^+\nu_\tau)$ are of similar magnitude for the $\tan\beta$ and $m_{H^+}$ regions of interest.

To incorporate helicity information throughout the whole process, it is
necessary to use a MC generator which can treat the full $2\to 6$ matrix element.
This is provided by MadGraph/MadEvent 4.1.10 \cite{Alwall&al:hep-ph:2007} for which we implemented the model (\ref{Eq:HiggsL}). Using this program has the additional advantage that the ME is generated without using the narrow width approximation. To verify the treatment of spin information, control samples were generated using the specialized $\ttbar$
generator TopReX 4.11 \cite{Slabospitsky&Sonnenschein:CPC:2002}. For the observables
we analyze, results from both programs were found to be in good agreement (after correcting the partial width $\Gamma(t\to bH^+)$ in TopReX).

From the $2\to6$ matrix element generated by MadEvent, full events are obtained by
applying Pythia 6.409 \cite{Sjostrand&al:JHEP:2006} for parton showering and hadronization.
To treat properly the spin information in tau decays, which was previously
demonstrated to be important in $H^\pm$ searches
\cite{Roy:PLB:1992,Raychaudhuri&Roy:PRD:1996}, Tauola \cite{Jadach&al:CPC:1993}
is invoked. The underlying event is modeled using the Pythia default "old"
model based on multiple parton-parton interactions. The default parameters of this
model are tuned to Tevatron minimum-bias data \cite{Field:hep-ph:2005} and provide reasonable estimates
for extrapolation to the LHC energy.

\subsection{Event Reconstruction}
\label{Sect:EvReco}

For jet reconstruction we use the FastJet implementation
\cite{Cacciari&Salam:PLB:2006} of the longitudinally invariant $k_\perp$
algorithm
\cite{Catani&al:NPB:1993,Ellis&Soper:PRD:1993,Dokshitzer&al:JHEP:1997} for hadron colliders. The jet clustering uses $E$-scheme recombination based on the distance measure
$d_{ij}=\mathrm{min}(k^2_{\perp i},k^2_{\perp j})\Delta R_{ij}^2/R^2$, with $\Delta
R=\sqrt{(\Delta\eta)^2+(\Delta\phi)^2}$ and $R^2=1$. We use the algorithm in \emph{exclusive} mode, meaning that particles are clustered with the beam when the beam-particle distance $d_{iB}$ is smaller than the distance $d_{ij}$ to any jet candidate. Not all particles will therefore end up in a jet. Furthermore, we only take particles with $|\eta|\leq 5$
into account to reflect the detector acceptance region. For the
minimum jet separation measure, above which no further clustering takes place,
the value $d_\mathrm{cut}=400$ (GeV)$^2$ is used. Choosing this value gives a jet multiplicity which peaks at the value expected from the matrix element.

We also implement a simplistic notion of flavor tagging where jets are tagged as $b$ jets or
$\tau$ jets by comparing to MC truth information. A candidate jet is tagged
whenever the distance $\Delta R$ to a true $b$ quark (or a $\tau$) is
less than $0.4$. In addition, for the jet to be tagged, it is required that it has $|\eta|\leq
2.5$. Apart from these criteria, no further efficiency factor is used in the
flavor tagging.
\begin{table}[b]
\centering
 \caption{Number of events generated at MC level, and events left after reconstruction. Integrated luminosities, corresponding to the generated event samples, are calculated using the NLO $\ttbar$ cross section. The luminosity varies due to the different branching ratios for the final state $\tau\nu_\tau jj$.}     
\begin{tabular}{ccccc}
   \hline
Model & &  \multicolumn{2}{c}{Number of events} &\\

  & &  Generated & Reconstructed & $\int\mathcal{L}$ (fb$^{-1}$) \\
         \hline
         SM           &                & $804\,926$    & $46\,524$ & $11.7$ \\[5pt]
         $t\bar{t}$ uncorr. &          & $937\,552$     & $54\,280$ & $13.7$ \\[5pt]
         2HDM (II)   & $\tan\beta=50$, $m_{H^+}=80$ GeV  & $925\,806$     & $59\,061$ & $3.84$ \\[5pt]
         2HDM (II)    & $\tan\beta=1$, $m_{H^+}=80$ GeV   & $926\,690$     & $59\,068$ & $7.77$ \\[5pt]
         \hline
\end{tabular}
\label{tab:Events}
\end{table}

Events are selected for analysis based on their overall topology. The
characteristic signature, which must be fulfilled by our signal events following
jet reconstruction, is the presence of
exactly two $b$ jets, one $\tau$ jet and at least two additional,
untagged, jets. For
this type of potential $\ttbar$ events, the
hadronically decaying $W$ is reconstructed by combining two light jets,
minimizing the mass-square difference $\Delta m_W^2=m_{jj}^2-m_W^2$. For $W$
candidates within a certain mass range $\Delta m_W\leq 10$~GeV of the true $W$
mass, further recombination with one of two $b$ jets is performed to reconstruct
a $t(\bar{t})$ candidate. On this candidate, a similar cut as for the
$W$ on $\Delta m_t^2=m_{jjb}^2-m_t^2$ is applied to asses the overall
goodness of the reconstruction. We keep events which have $\Delta m_t<
15$~GeV. For the surviving events, one side has then been
fully reconstructed. The remaining $b$ jet, not used in the top
reconstruction, can therefore be associated with the $t\to bH^\pm/W^\pm\to b\tau^\pm\nu_\tau$ decay.

To show the statistical efficiency, we give in Table~\ref{tab:Events} the total number of MC events generated for each model. Since the branching ratios to reach the requested final state are largely different for the SM and the 2HDM, this nearly constant number of events actually corresponds to quite different integrated luminosities, as also given in the table. Note that, for these low Higgs masses, the luminosity corresponding to this number of events is highest for the SM sample. Table~\ref{tab:Events} also lists the total number of events left in each case after reconstruction. These numbers indicate the level of statistical uncertainty in the jet-level distributions we show below.

Due to the presence of at least \emph{two}
neutrinos in the final state from the decay of the $H^\pm$ (or $W^\pm$) into
$\tau^\pm\nu_\tau$, the longitudinal momentum of the $t$($\bar{t}$) on that side
of the event cannot be reconstructed. Thus its rest system is not accessible,
neither is the overall CM frame. This fact effectively prevents the direct
experimental use of distributions such as (\ref{Eq:ddiff2}) or (\ref{Eq:diffct}) to establish the presence of a $t\to bH^+$ channel. This is in contrast to the SM
case where the semi-leptonic ($e$,$\mu$) channels could be very useful in establishing the existence of $\ttbar$ spin correlations. In the following sections, we will focus on how to address this somewhat discouraging situation.

\subsection{Distributions in $\Delta\phi$}
We will first investigate if there are interesting observables defined directly in the laboratory frame.
An early study \cite{Barger&al:IJMP:1989} established that
distributions in azimuthal angle $\Delta\phi(l^+l^-)$, with the $l^+l^-$
originating from two leptonic decays of a $\ttbar$ pair, are sensitive to
$\ttbar$ spin correlations. These results were recently confirmed in a NLO
shower MC study \cite{Frixione&al:JHEP:2007}, where also other spin-sensitive observables
defined directly from lab quantities were considered. Their conclusion was that,
for all the variables included in the study (see \cite{Frixione&al:JHEP:2007}
for the complete list), the spin effects are only visible in the distribution
$\Delta\phi(l^+ l^-)$.

\begin{figure}
\begin{centering}
\subfigure{
   \label{Fig:A}
   \includegraphics[width=0.45\columnwidth,keepaspectratio]{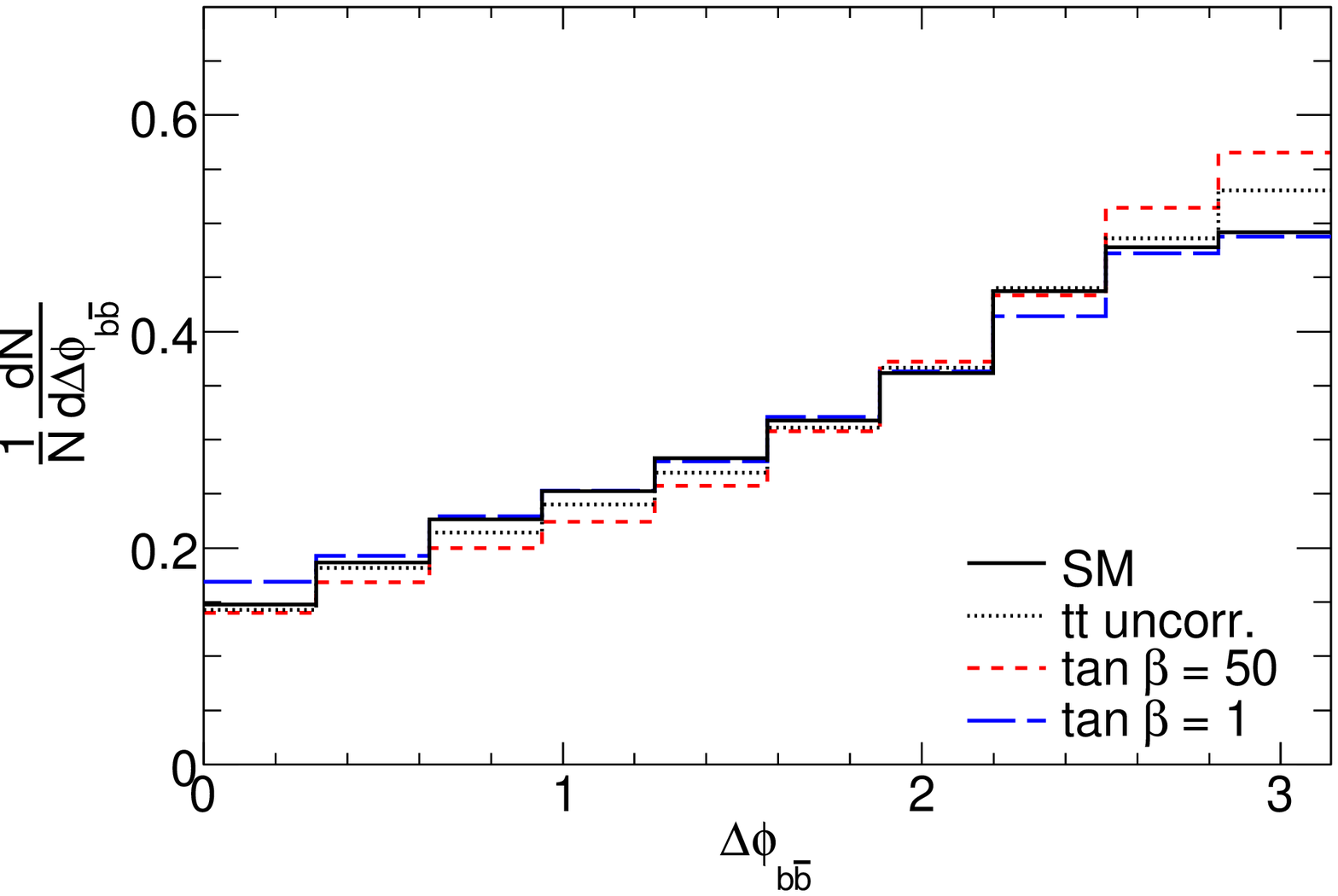}

}
\subfigure{
   \label{Fig:B}
   \includegraphics[width=0.45\columnwidth,keepaspectratio]{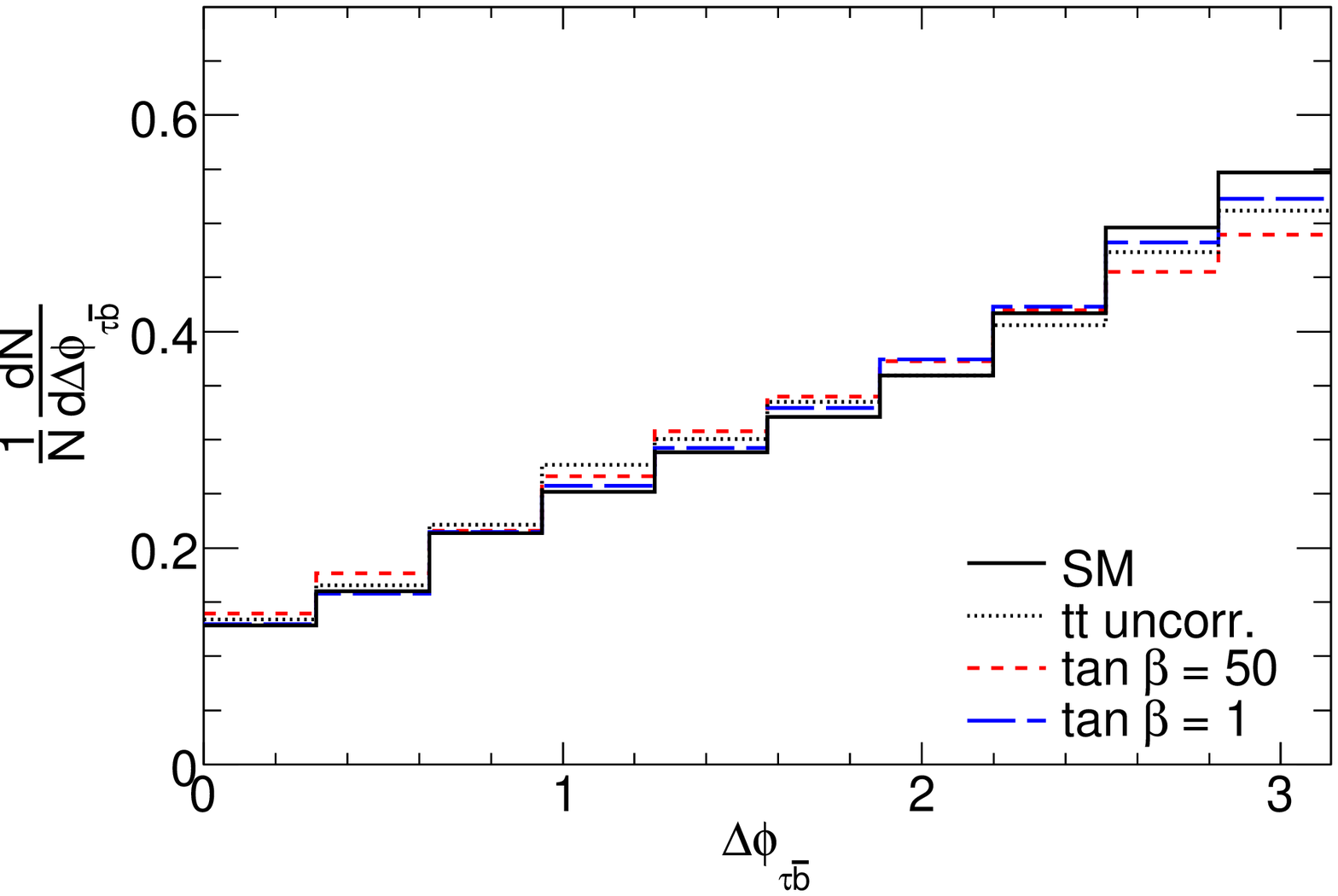}
}
\caption{Azimuthal angle between $b$ jet from $W$ decay and opposite
side $b$- or $\tau$ jet. The results are shown for the SM $\ttbar$ (black), or $t\to bH^+$ decay with $\tan\beta=50$ (short-dashed red) and $\tan\beta=1$ (long-dashed blue). For comparison, the distribution obtained in the SM without spin correlations is shown as a dotted black line. The mass $m_{H^+}=80$~GeV was used.}
\label{Fig:dphi_lab}
\end{centering}
\end{figure}

Since we require hadronic reconstruction of a $W$ boson and a top quark, we are effectively limited to studying the two distributions $\Delta\phi(b\bar{b})$ and $\Delta\phi(\tau \bar{b})$. From what we know about the spin analyzing efficiencies,
 we could expect that, at least for the Higgs case, these variables could show some effects which did not show in the SM study.
The resulting azimuthal distributions are shown in Figure~\ref{Fig:dphi_lab}. In addition to showing the SM and the 2HDM (II) for $\tan\beta=50$, and $\tan\beta=1$, we show also this time the SM \emph{without} any spin correlations to illustrate that this distribution is not a flat line. From the figure, in the $b\bar{b}$ distribution, it can be seen again that the low $\tan\beta$ case is more SM-like and that the high $\tan\beta$ case gives the most significant deviation from the SM distribution. In the $\tau b$ distribution, the curves are hardly separable. It therefore remains to be seen if these variables could be used in any way to discriminate between the SM and new physics contributions.

Finally we note that these distributions are quite sensitive to changes in the Higgs mass. Larger values of $\Delta\phi$ are favored with increasing $m_{H^+}$. The kinematic effects must therefore be under good control if the $\Delta\phi$ distributions should be used to determine $\tan\beta$.

\subsection{Brandenburg observables}
We now turn our attention from the laboratory frame observables to what can be done with half the event fully reconstructed. This has been discussed by Brandenburg who introduced  \cite{Brandenburg:PLB:1996} the observables
\begin{equation}
\begin{aligned}
\mathcal{O}_1&=\mathbf{k_1^*}\cdot\mathbf{\tilde{k}_2}\\
\mathcal{O}_2&=(\mathbf{k_1^*}\cdot
\mathbf{\hat{z}})(\mathbf{\tilde{k}_2\cdot\hat{z}})\\
\mathcal{O}_3&=(\mathbf{k_1^*\cdot p})(\mathbf{\tilde{k}_2\cdot p}),
\end{aligned}
\end{equation}
where an asterisk denotes momenta in the rest frame of the parent top quark. The beam direction is given by $\mathbf{\hat{z}}$,
and $\mathbf{\tilde{k}_2}$ is the laboratory momentum of the lepton from the top decay not associated with $\mathbf{k_1}$. Each
of the observables $\mathcal{O}_1$ -- $\mathcal{O}_3$ are cleverly constructed to have mean value strictly
equal to zero in the absence of $\ttbar$ spin correlations.
\begin{table}[b]
\centering
 \caption{Mean value of the Brandenburg observables in percent. Evaluated for
SM ($W^\pm W^\mp$) and charged Higgs ($W^\pm H^\mp)$ decay of $\ttbar$ with $m_{H^+}=80$ GeV.}     
\begin{tabular}{ccr@{.}lr@{.}lr@{.}lc}
   \hline
Model & &  \multicolumn{6}{c}{Observable}\\

  & &  \multicolumn{2}{c}{$\Ave{\mathcal{O}_1}$}
   & \multicolumn{2}{c}{$\Ave{\mathcal{O}_2}$}
   & \multicolumn{2}{c}{$\Ave{\mathcal{O}_3}$} & \\
         \hline
         SM      &       & -2&08  & -0&25 & -0&67 & \% \\[5pt]
         2HDM (II)&$\tan\beta=1$  & -1&06 & -0&11 & -0&31  & \% \\[5pt]
         2HDM (II) &$\tan\beta=50$ &  0&86 &  0&04 &  0&24  & \% \\[5pt]
         \hline
\end{tabular}
\label{tab:Brandenburg}
\end{table}
We first evaluated $\mathcal{O}_1$ -- $\mathcal{O}_3$ at parton-level, similarly to what is done in \cite{Brandenburg:PLB:1996}, but also including the decay through $H^\pm$. The resulting mean values obtained from the distributions in these observables are presented in Table \ref{tab:Brandenburg}. Their standard deviations are $\Delta\mathcal{O}_1=0.58$, $\Delta\mathcal{O}_2=0.42$ and $\Delta\mathcal{O}_3=0.39$ for all three cases. Comparing our SM values with those given in the original study, we find that there is an overall agreement. For the Higgs case, we observe the same sign differences between the low and high $\tan\beta$ regimes as previously seen for the correlation plots. We also note that the Higgs values are generally smaller than their SM counterparts.

Unfortunately, when evaluated at the jet level, including the cuts described above, we find that the expectation values for all three Brandenburg observables are consistent with zero.

\subsection{Transverse correlations}
\begin{figure}
\begin{centering}
\subfigure{
   \includegraphics[width=0.45\columnwidth,keepaspectratio]{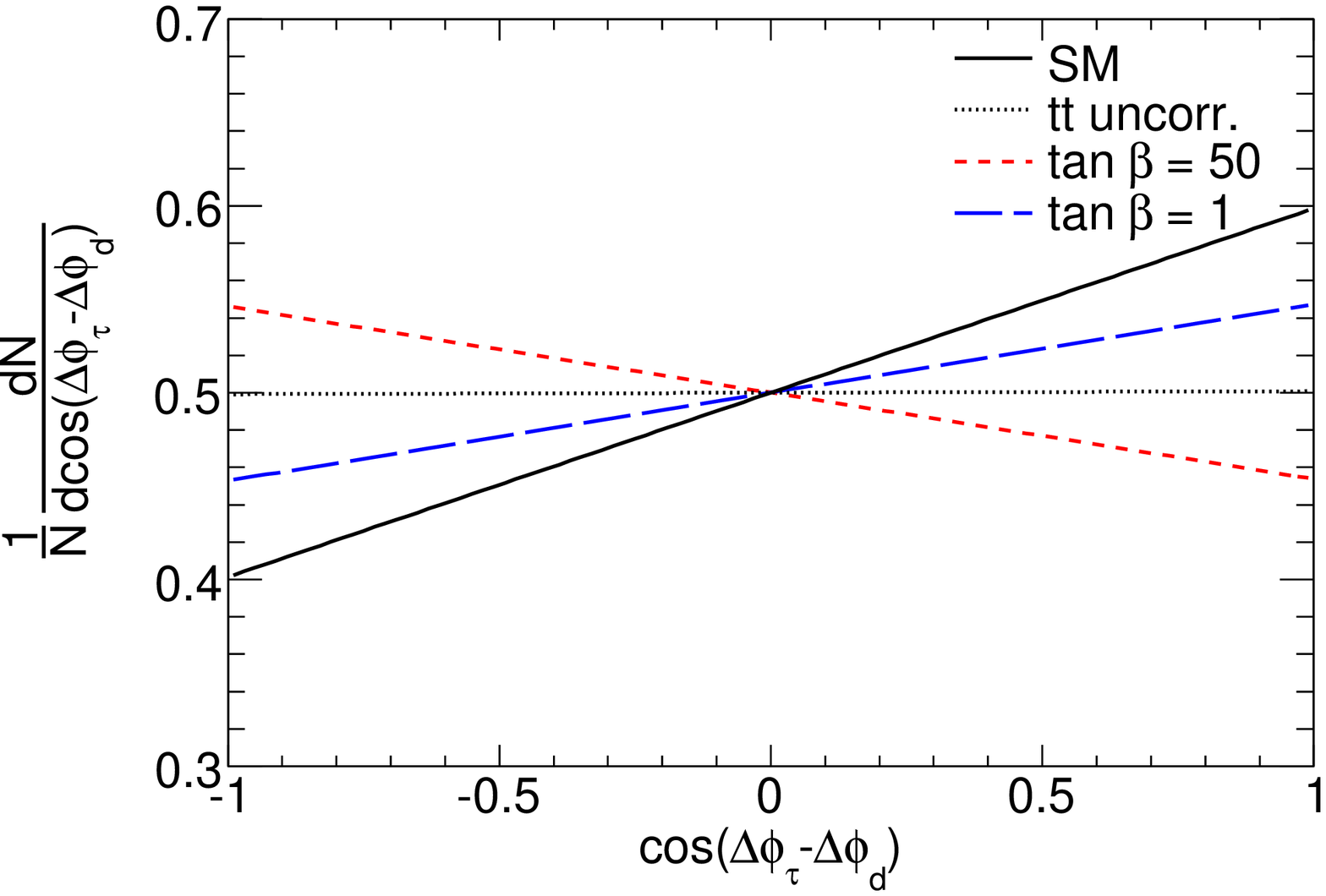}

}
\subfigure{
   \includegraphics[width=0.45\columnwidth,keepaspectratio]{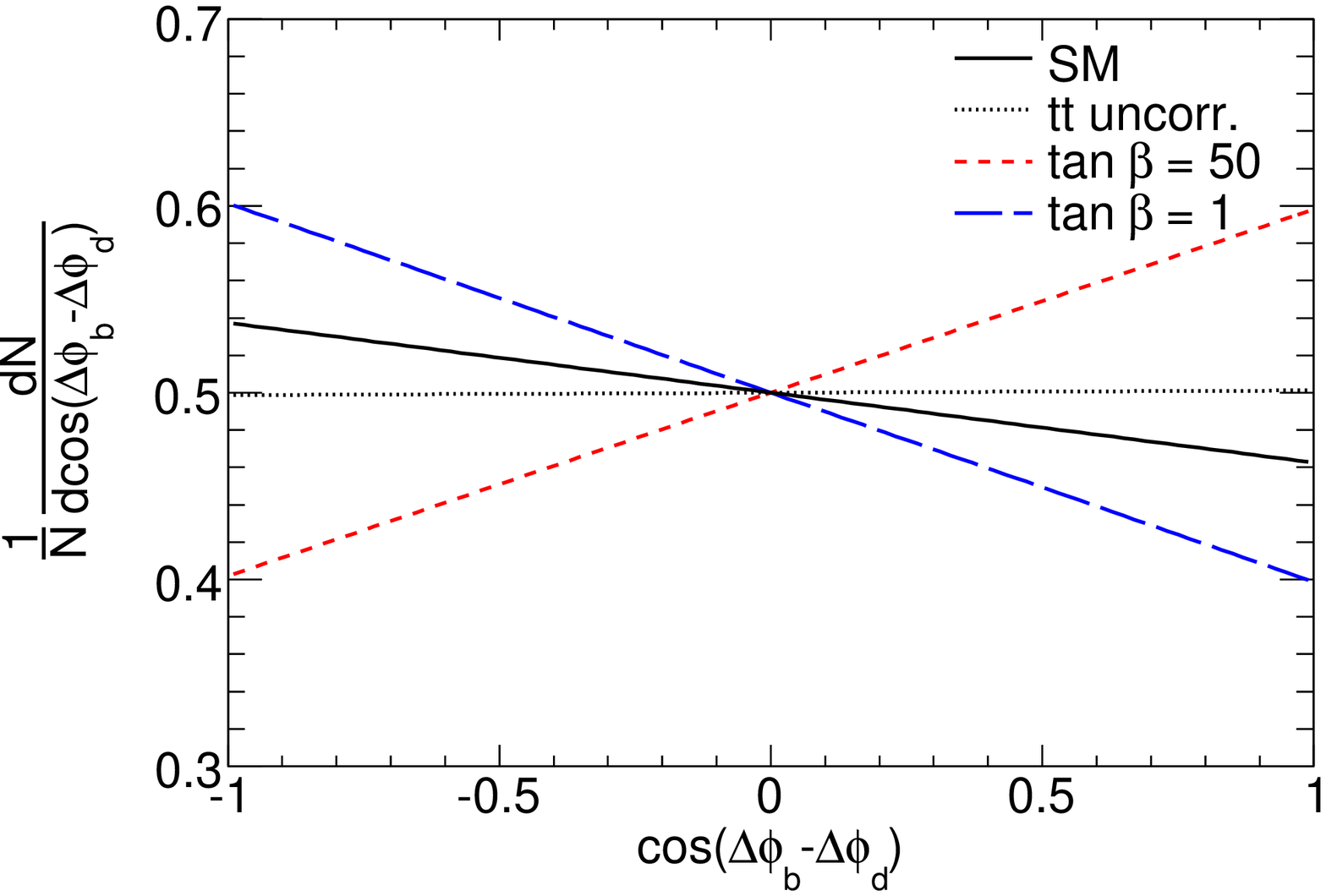}
}
\subfigure{
   \includegraphics[width=0.45\columnwidth,keepaspectratio]{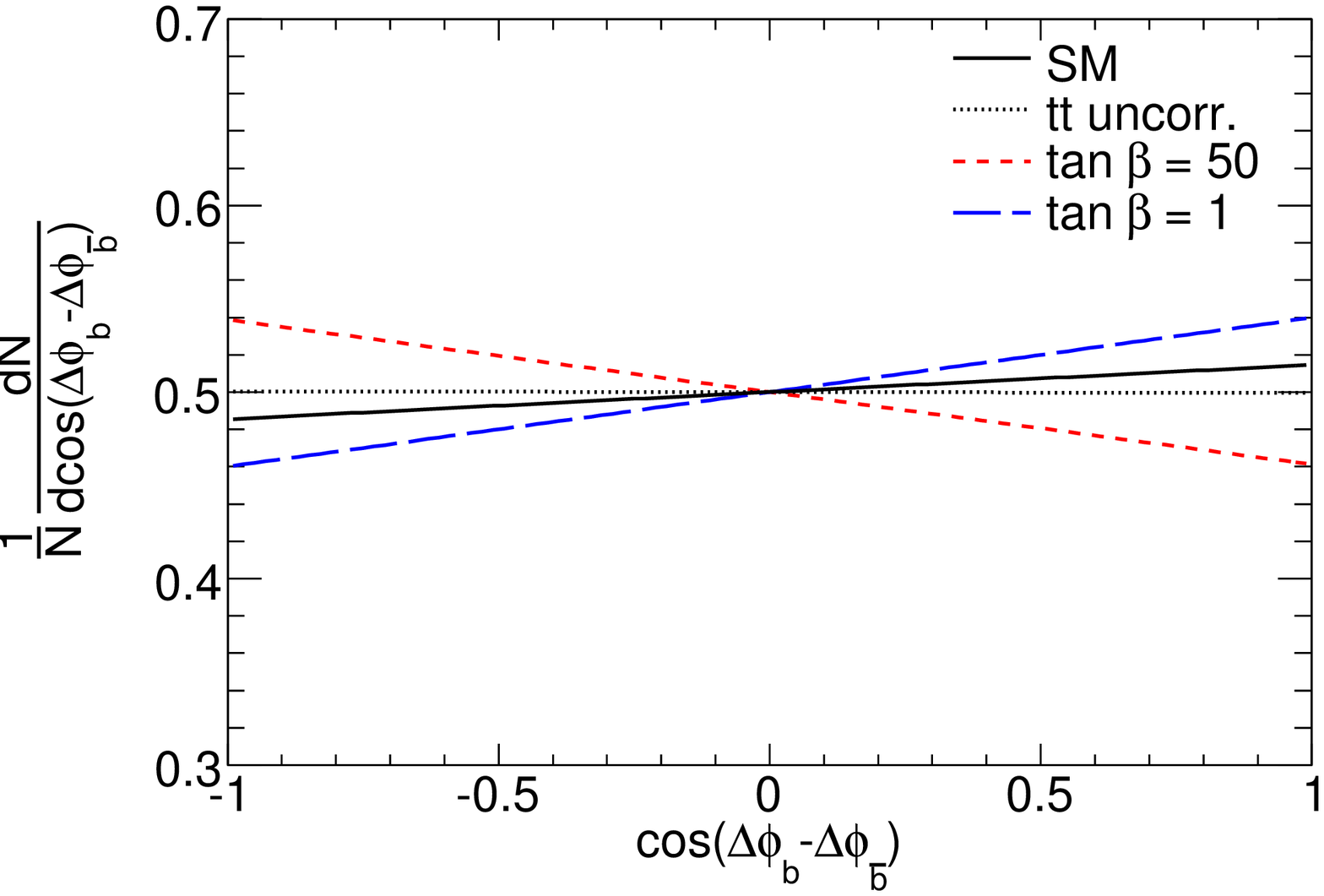}

}
\subfigure{
   \includegraphics[width=0.45\columnwidth,keepaspectratio]{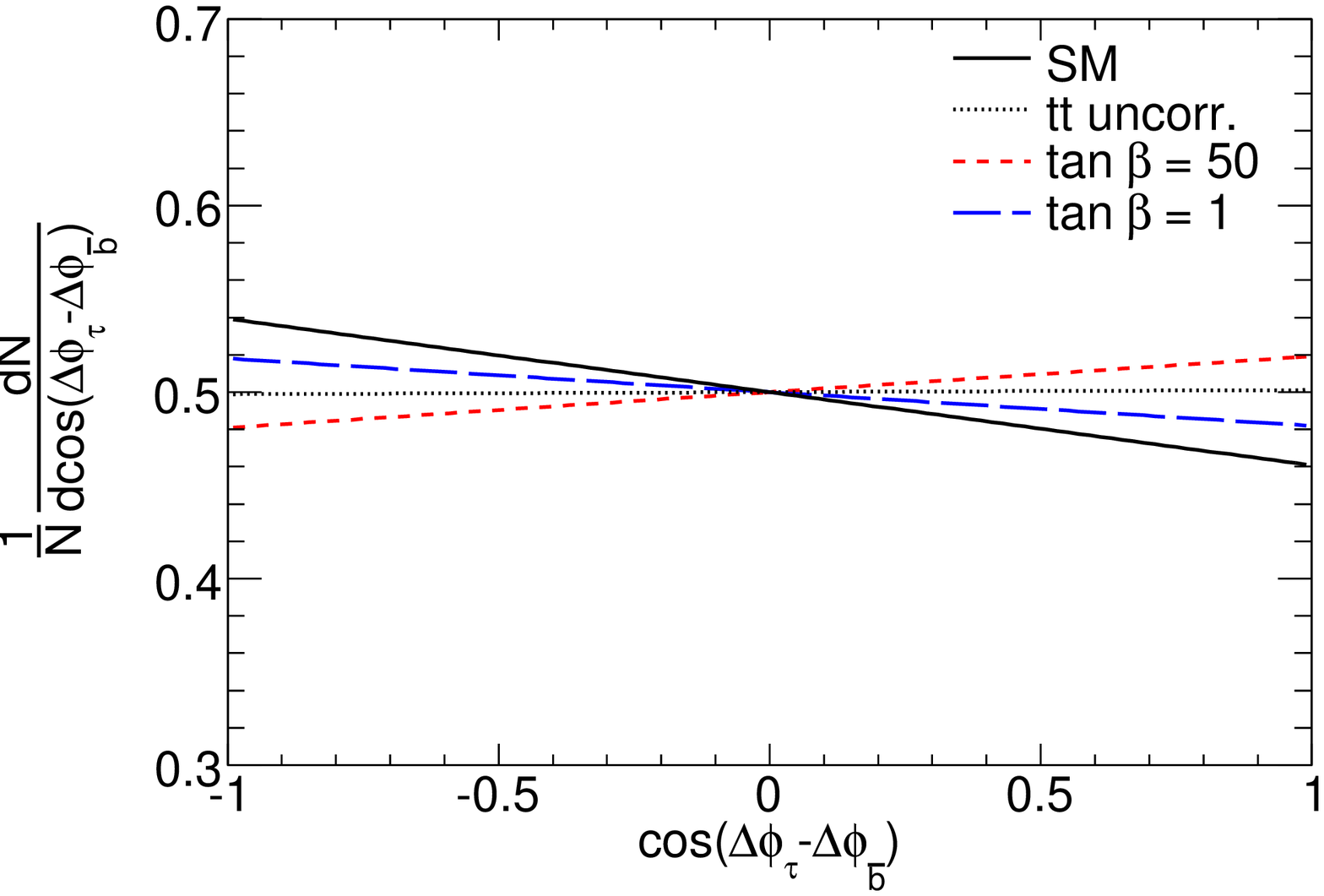}
}
\caption{Distributions at matrix element level of $\cos(\Delta\phi_i-\Delta\phi_j)$ for different final state particles $(i,j)$. Dotted line show results in the SM \emph{without} spin correlations. SM results with correlations are shown in solid black. The 2HDM (II) is shown for $\tan\beta=1$ (dashed blue), and $\tan\beta=50$ (dotted red). All results were obtained for $m_{H^+}=80$ GeV.}
\label{Fig:dphi_ME}
\end{centering}
\end{figure}

As an alternative to the full reconstruction of the rest frames for $t$ and
$\bar{t}$, and the half reconstruction performed when calculating the Brandenburg observables, we consider transverse reconstruction. This is possible in the hadronic channel where one top quark is reconstructed using the methods discussed above, while the transverse momentum of the other top can be obtained from the sum
$p_{\perp,t}=p_{\perp,b}+p_{\perp,\tau}+p_{\perp,\mathrm{miss}}$. The
sum involves the $b$ jet not used for top reconstruction in the
first step.

Following the reconstruction, we perform boosts of particles $(i,j)$ into the
\emph{transverse} rest frames of $t(\bar{t})$, and consider distributions in this frame of azimuthal angles $(\Delta\phi_i,\Delta\phi_j)$ to the $p_\perp$ axes of
$t(\bar{t})$. It is then possible to form both doubly differential
distributions similar to (\ref{Eq:ddiff2}), or one-dimensional distributions
similar to (\ref{Eq:diffct}). To obtain a distribution which like (\ref{Eq:diffct}) is robust with respect to cuts, we have looked specifically at distributions in $\Delta\phi_i-\Delta\phi_j$. In analogy with the construction of CM variables, we expect a distribution of the form
\begin{equation}
\label{Eq:diffdphi}
\frac{1}{N}\frac{\mathrm{d}N}{\mathrm{d}\cos(\Delta\phi_i-\Delta\phi_j)}=\frac{1}{2}\Bigl[1+\mathcal{D'}\alpha_i\alpha_j\cos(\Delta\phi_i-\Delta\phi_j)\Bigr].
\end{equation}
This relation involves yet another spin correlation coefficient $\mathcal{D'}$. To illustrate how similar this new observable is to $\cos\theta_{ij}$, which was previously shown in Figure~\ref{Fig:cost}, Figure~\ref{Fig:dphi_ME} shows the matrix element results on $\cos(\Delta\phi_i-\Delta\phi_j)$ for the same particle combinations. The results are presented for the SM, and for the charged Higgs decay with $\tan\beta=50$ and $\tan\beta=1$. From comparing the two figures, we note that $\mathcal{D'}\simeq0.9\mathcal{D}$, signaling only a slight loss of correlation going to the transverse projection.

\begin{figure}
\begin{centering}
\subfigure{
   \includegraphics[width=0.45\columnwidth,keepaspectratio]{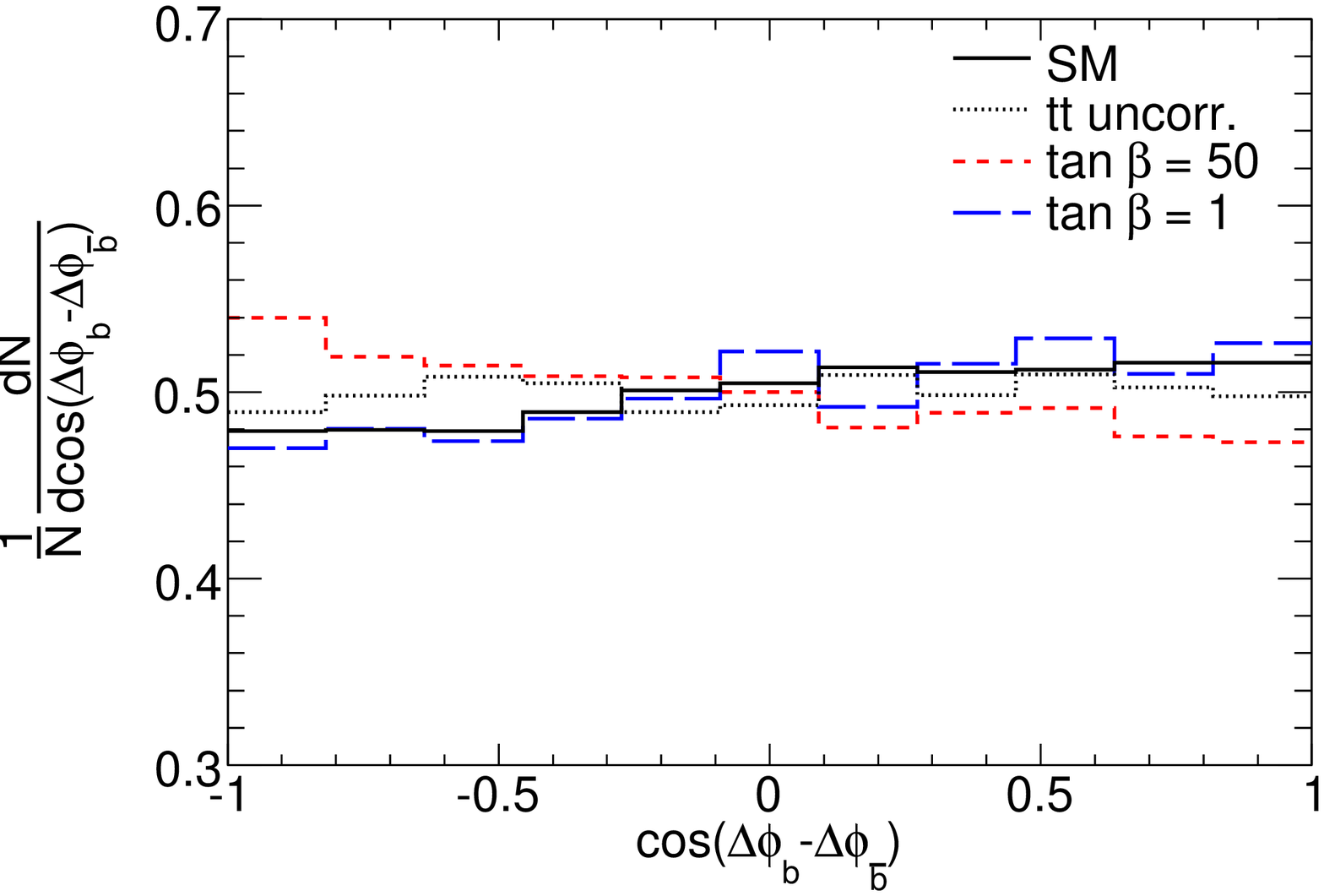}

}
\subfigure{
   \includegraphics[width=0.45\columnwidth,keepaspectratio]{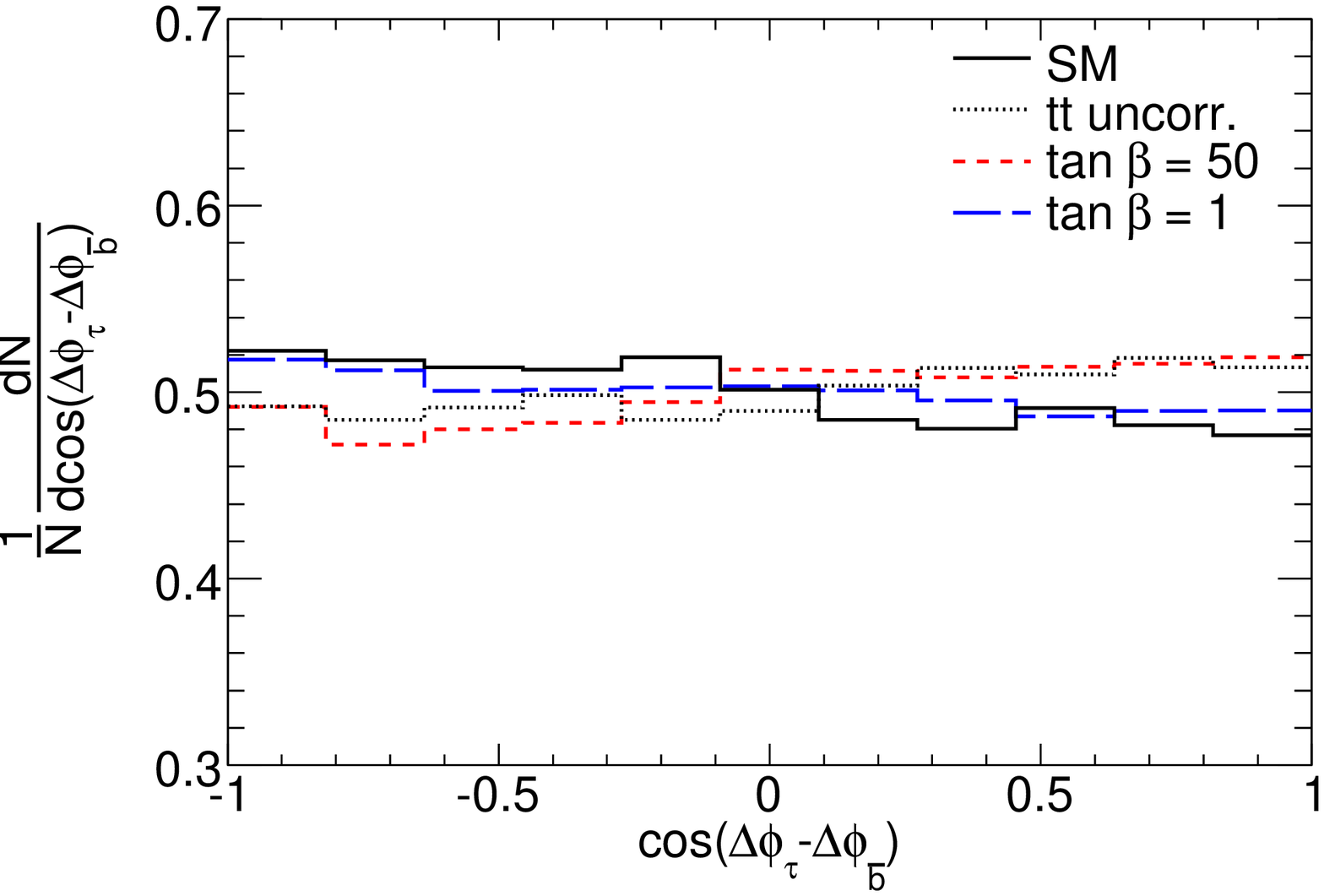}
}
\caption{Distributions at jet level of $\cos(\Delta\phi_i-\Delta\phi_j)$ for different final state particles $(i,j)$. The different contours correspond to the SM with (solid black), and without (black dotted), $\ttbar$ spin correlations. There is also the 2HDM (II) for $\tan\beta=1$ (dashed blue), and for $\tan\beta=50$ (dotted red). All results were obtained for $m_{H^+}=80$ GeV.}
\label{Fig:dphi_jet}
\end{centering}
\end{figure}

These transverse correlations also maintain their properties when going to the level of jets, as demonstrated in Figure~\ref{Fig:dphi_jet}. That only two diagrams are shown in Figure~\ref{Fig:dphi_jet}, and not even the ones from Figure~\ref{Fig:dphi_ME} with best separation between the different cases, is because the efficiency in experimentally separating a $d$-type jet from a $u$-type jet is only $61\%$ \cite{Mahlon&Parke:PRD:1996}. Since this separation step is required in order to use the jets from the hadronic $W$ decay in the analysis, we exclude the $(\tau,d)$ and $(b,d)$ combinations, which both require this information. We remind the reader that the distribution $(H^+,\bar{b})$ contains the same information as that for $(b,\bar{b})$. This is true also at jet level, since the momenta of $H^+$ and $b$ are still anti-parallel in the reconstructed top rest frame.
Figure~\ref{Fig:dphi_jet} shows good separation between the SM and the 2HDM (II) for $\tan\beta=50$, while the small $\tan\beta$ case is more difficult to separate from the SM. Since this small separation is partly intrinsic, as can be seen from the lower panels in Figure~\ref{Fig:dphi_ME}, these variables shows most promise in the high $\tan\beta$ regime.

We have studied the dependence of the distributions in Figure~\ref{Fig:dphi_jet} on $m_{H^+}$. From their $\alpha$ dependence, the coupling sensitivity in the $\tau\bar{b}$ correlation is expected to degrade for higher $m_{H^+}$, while it should not change for $b\bar{b}$. This is indeed what is observed. We also find additional distortion in both distributions when $m_{H^+}\gtrsim 130$ GeV. For $b\bar{b}$, values $\cos(\Delta\phi_b-\Delta\phi_{\bar{b}})<0$ are favored, while for the $\tau\bar{b}$ distribution instead $\cos(\Delta\phi_\tau-\Delta\phi_{\bar{b}})>0$ shows a surplus of events. However, due to the small values of $\alpha_\tau$ for $H^\pm$, the $\tau\bar{b}$ distribution is hardly interesting for such masses. This additional angular dependence, which is not observed at parton level, we attribute to the $b$ jet reconstruction. For higher $m_{H^+}$, the $b$ quarks from $t\to bH^+$ becomes softer, and eventually falls below the jet measure $d_\mathrm{cut}$ in $p_\perp$. These events should normally not pass for further analysis, if not the underlying event provided the necessary energy. When these jets are dominated by the underlying event, their directions are smeared, causing the $t$ rest frame to be poorly reconstructed. The result is an additional distortion in the angular distributions compared to the ME results.

\section{Summary and conclusions}
\label{Sect:Conclusions}
Top quarks produced in pairs at the LHC should experience strong correlations
among their spin projections in the helicity basis. This entanglement makes it a
favorable system for studying the Lorentz structure of the couplings involved in
top quark production and decay. 
We have considered here the effects on spin dependent angular observables of new
physics in top quark decays. Specifically, we have discussed top decay through a
charged Higgs boson.

Within a general CP-conserving 2HDM, we obtain results on the spin analyzing
coefficients. They determine the sensitivity of a given decay product to the
spin projection of the decaying particle. Unlike the SM decay $t\to bW^+\to
bl^+\nu_l$, where the charged lepton (or $d$ quark jet) is
the most powerful spin analyzer, the scalar case has the associated $b$ quark, or
the $H^\pm$ momentum itself, as the most efficient spin analyzers. As for the lepton in the SM case, this efficiency can reach unity. In the 2HDM (II), this is found to be the case for very large (and very small) values of $\tan\beta$. At the level of matrix elements, the modification to the distributions of certain
angular observables, induced by the presence of an $H^\pm$, could therefore be similar in magnitude to the SM spin effects themselves. Whether this is actually
the case depends on the parameters of the 2HDM. Since all spin analyzing coefficients
associated with $t\to bH^\pm\to b\tau^\pm\nu_\tau$ depend universally on
the Lorentz structure of the $tbH^+$ vertex, an angular analysis of the decay products 
could provide an additional handle on the $H^\pm$ couplings.
This includes the possibility of determining the effective value of
$\tan\beta$ in the MSSM, since we find that $\tan\beta$ enhanced SUSY corrections to the spin correlation observables are small.

The preference of the $H^\pm$ to decay into $\tau^\pm\nu_\tau$ prevents
reconstruction of longitudinal momenta. Thus the $t(\bar{t})$ rest frame spin analysis, which can be performed in the SM, cannot be directly applied to the new physics case. Encouraged by the significant spin effects observed at matrix element level, we have therefore studied several longitudinally boost-invariant observables. 

We find that the distribution in azimuthal angle between the $b$ jet associated with $t\to bH^+$, and the $b$ jet from the opposite side $t\to bW$ decay, shows some sensitivity to the Lorentz structure of the $H^\pm$ coupling. Although the effects on this observable are small, they are directly measurable and therefore shows promise for further experimental study. 

Even more promising are $(\Delta\phi_i,\Delta\phi_j)$ correlations in the transverse rest frames of $t(\bar{t})$. This observable is also directly measurable, and it is found to be robust with respect to phase-space cuts. The highest sensitivity is also here obtained for the $b\bar{b}$ correlation. This distribution should be particularly useful in the highly interesting case with large $\tan\beta$, as it lifts the degeneracy between large and small $\tan\beta$ values present in the inclusive $\mathcal{BR}(t\to bH^+)$ measurement. It should be investigated if this observable could be useful also for studying $\ttbar$ spin effects within the SM. In that case, it should be possible to use $lb$ distributions where $l=e,\mu$.
Another question is if the correlations in $bb$ or $lb$ obtained from the semi-leptonic $\ttbar$ decay in the SM could be used as control samples. 

As illustrated by our MC samples, statistics is not the limiting factor in an analysis of this type due to the large cross section for $\ttbar$ production at the LHC. We therefore conclude that further investigation of spin effects in top physics beyond the Standard Model should be worthwhile.

\section*{Acknowledgments}
The authors would like to thank Elias Coniavitis and Nazila Mahmoudi for interesting and useful discussions.

\appendix
\section{Massive Spinor formalism}
\label{App:Spin}
In this appendix, we discuss the formalism used to describe polarized massive fermions. Throughout
the discussion, we try to adhere to the conventions of \cite{Haber:hep-ph:1994}
which contains an excellent introduction to spin formalism.

For massive spin-$1/2$ particles, the projection relations 
\begin{equation}
\begin{aligned}
\label{Eq:proj_sum}
\sum_{\lambda=\pm\frac{1}{2}}u(p,\lambda)\bar{u}(p,\lambda)&=\slashed{p}+m\\
\sum_{\lambda=\pm\frac{1}{2}}v(p,\lambda)\bar{v}(p,\lambda)&=\slashed{p}-m
\end{aligned}
\end{equation}
obtained by summing over the possible helicity states $\lambda$ are very
well-known, as are the more general projector relations 
\begin{equation}
\begin{aligned}
\label{Eq:proj_spin}
u(p,\lambda)\bar{u}(p,\lambda)&=\frac{1}{2}\left(1+\gamma_5\slashed{s}
\right)\left(\slashed{p}+m\right)\\
v(p,\lambda)\bar{v}(p,\lambda)&=\frac{1}{2}\left(1+\gamma_5\slashed{s}
\right)\left(\slashed{p}-m\right)
\end{aligned}
\end{equation}
with a time-like spin four-vector defined as
\begin{equation}
\label{Eq:s_mu}
s^\mu=2\lambda\left(\frac{\left|\mathbf{p}\right|}{m},\frac{E}{m}
\mathbf{\hat{p}}\right).
\end{equation}
In the rest frame of the massive particle $s^\mu=2\lambda(0,\hat{\mathbf{p}})$,
whereas in the high-energy limit $s^\mu=2\lambda p^\mu/m$.
As a further generalization of the projector technique, Bouchiat and Michel
\cite{Bouchiat&Michel:NP:1958} introduced a projection relation for the more
general product of two Dirac spinors of mass $m$ with arbitrary helicities. For
this we need to introduce three spin vectors $s_\mu^a$ that are mutually
orthonormal, and in addition orthogonal to $p/m$ in the sense:
\begin{equation}
\begin{aligned}
p\cdot s^a &= 0\\
s^a\cdot s^b &= -\delta^{ab}\\
s_\mu^a s_\nu^a &= -g_{\mu\nu}+\frac{p_\mu p_\nu}{m^2}\ .
\end{aligned}
\end{equation}
The most direct choice for an explicit spin basis of this type is perhaps, in a
coordinate system where $\hat{\mathbf{p}}$ is in the positive $z$-direction, to
use 
\begin{equation}
\begin{aligned}
\label{Eq:sbasis}
s^{1\mu}&=(0,\mathbf{\hat{x}})\\
s^{2\mu}&=(0,\mathbf{\hat{y}})\\
s^{3\mu}&=\left(\frac{|\mathbf{p}|}{m},\frac{E}{m}\mathbf{\hat{p}}\right).
\end{aligned}
\end{equation}
With this choice, $s^{3\mu}$ differs from the positive helicity spin vector
(\ref{Eq:s_mu}) only by a factor $2\lambda$. The helicity spinors can be shown to satisfy
\begin{equation}
\begin{aligned}
\gamma_5\slashed{s}^a u(p,\lambda')&=\sigma_{\lambda\lambda'}^au(p,\lambda)\\
\gamma_5\slashed{s}^a v(p,\lambda')&=\sigma_{\lambda'\lambda}^av(p,\lambda)
\end{aligned}
\end{equation}
where the rhs is summed over $\lambda$, and the $\sigma^a$ are Pauli
matrices where the first (second) row and column correspond to $\lambda=1/2$
($\lambda=-1/2$). Thus $\sigma^3=2\lambda I$.
From these identities, the Bouchiat-Michel relations
\begin{equation}
\begin{aligned}
\label{Eq:BM}
u(p,\lambda')\bar{u}(p,\lambda) &=
\frac{1}{2}\left(\delta_{\lambda\lambda'}+\gamma^5\slashed{s}^a\sigma^a_{
\lambda\lambda'}\right)\left(\slashed{p}+m\right) \\
v(p,\lambda')\bar{v}(p,\lambda) &=
\frac{1}{2}\left(\delta_{\lambda\lambda'}+\gamma^5\slashed{s}^a\sigma^a_{
\lambda'\lambda}\right)\left(\slashed{p}-m\right).
\end{aligned}
\end{equation}
follow. Letting $\lambda'=\lambda$ (no sum over $\lambda$), and noting that
$s^i_{\lambda\lambda}\neq 0$ only for $i=3$ with the spin basis
(\ref{Eq:sbasis}), the relations (\ref{Eq:proj_spin}) are obtained
as a special case of the Bouchiat-Michel formulae.

\bibliographystyle{JHEP}
\bibliography{ttspin}

\end{document}